\documentclass[aps,prc,twocolumn,superscriptaddress,nofootinbib]{revtex4-1}

\usepackage{graphicx}
\usepackage{hyperref}

\begin{document}


\title{Proton Capture on $^{17}$O and its astrophysical implications}


\author{A.~Kontos}
\altaffiliation[Present address: ]{National Superconducting Cyclotron Laboratory, Michigan State University, East Lansing, MI 48824, USA}
\email[E-mail address: ]{kontos@nscl.msu.edu}
\author{J.~G\"orres}
\author{A.~Best}
\altaffiliation[Present address: ]{Lawrence Berkeley National Laboratory, Berkeley, CA 94720, USA}
\author{M.~Couder}
\author{R.~deBoer}
\affiliation{Department of Physics, University of Notre Dame, Notre Dame, Indiana 46556, USA}
\affiliation{The Joint Institute for Nuclear Astrophysics, University of Notre Dame, Notre Dame, IN 46556, USA}

\author{G.~Imbriani}
\affiliation{The Joint Institute for Nuclear Astrophysics, University of Notre Dame, Notre Dame, IN 46556, USA}
\affiliation{Dipartimento di Scienze Fisiche, Universit\`a Federico II, and INFN Sezione di Napoli, Naples, Italy}

\author{Q.~Li}
\author{D.~Robertson}
\affiliation{Department of Physics, University of Notre Dame, Notre Dame, Indiana 46556, USA}
\affiliation{The Joint Institute for Nuclear Astrophysics, University of Notre Dame, Notre Dame, IN 46556, USA}

\author{D.~Sch\"urmann}
\altaffiliation[Present address: ]{Dipartimento di Scienze Fisiche, Universit\`a Federico II, Naples, Italy}
\affiliation{The Joint Institute for Nuclear Astrophysics, University of Notre Dame, Notre Dame, IN 46556, USA}

\author{E.~Stech}
\author{E.~Uberseder}
\author{M.~Wiescher}
\affiliation{Department of Physics, University of Notre Dame, Notre Dame, Indiana 46556, USA}
\affiliation{The Joint Institute for Nuclear Astrophysics, University of Notre Dame, Notre Dame, IN 46556, USA}



\date{\today}

\begin{abstract} 
\begin{description} 
\item[Background] The reaction $^{17}$O$(p,\gamma)^{18}$F influences hydrogen-burning nucleosynthesis in several stellar sites, such as red giants, asymptotic giant branch (AGB) stars, massive stars and classical novae. In the relevant temperature range for these environments ($T_{9}=0.01$-$0.4$), the main contributions to the rate of this reaction are the direct capture process, two low lying narrow resonances ($E_{r}=65.1$ and $183$ keV) and the low-energy tails of two broad resonances ($E_{r}=557$ and $677$ keV).
\item[Purpose] Previous measurements and calculations give contradictory results for the direct capture contribution which in turn increases the uncertainty of the reaction rate. In addition, very few published cross section data exist for the high energy region that might affect the interpretation of the direct capture and the contributions of the broad resonances in the lower energy range. This work aims to address these issues. 
\item[Method]  The reaction cross section was measured in a wide proton energy range ($E_{c.m.}=345$ - $1700$ keV) and at several angles ($\theta_{lab}=0^{\circ},45^{\circ},90^{\circ},135^{\circ}$). The observed primary $\gamma$-transitions were used as input in an $R$-matrix code in order to obtain the contribution of the direct capture and the two broad resonances to the low-energy region.
\item[Results] The extrapolated S-factor from the present data is in good agreement with the existing literature data in the low-energy region. A new reaction rate was calculated from the combined results of this work and literature S-factor determinations. Resonance strengths and branchings are reported for several $^{18}$F states.
\item[Conclusions] We were able to extrapolate the astrophysical S-factor of the reaction $^{17}$O$(p,\gamma)^{18}$F at low energies from cross section data taken at higher energies. No significant changes in the nucleosynthesis are expected from the newly calculated reaction rate.
\end{description} 
\end{abstract}


\maketitle

\section{Introduction}

Proton induced reactions on $^{17}$O nuclei take place in the hydrogen burning shells of red giants, asymptotic giant branch (AGB) stars, and the cores of massive stars, at temperatures around T $=0.03-0.1$ GK. At these conditions hydrogen burning is dominated by the CNO cycles~\cite{iliadisstars}, namely the reaction sequences $^{12}$C$(p,\gamma)^{13}$N$(\beta^+,\nu)^{13}$C$(p,\gamma)^{14}$N$(p,\gamma)^{15}$O$(\beta^+,\nu)$ $^{15}$N$(p,\alpha)^{12}$C (CNO-I), $^{15}$N$(p,\gamma)^{16}$O$(p,\gamma)^{17}$F$(\beta^+,\nu)^{17}$O $(p,\alpha)^{14}$N (CNO-II), $^{17}$O$(p,\gamma)^{18}$F$(\beta^+,\nu)^{18}$O$(p,\alpha)^{15}$N (CNO-III), and $^{18}$O$(p,\gamma)^{19}$F$(p,\alpha)^{16}$O (CNO-IV). Of particular interest is the branching between the two reactions  $^{17}$O$(p,\gamma)^{18}$F and $^{17}$O$(p,\alpha)^{14}$N that determines the leakage towards the third and fourth CNO cycles. The reaction rates of these two proton induced reactions determines the branching ratio which in turn affects the nucleosynthesis and the abundance ratio of the rare $^{17}$O and $^{18}$O isotopes at different environmental conditions of the burning site. Comparison with observed abundance distributions of the oxygen isotopes will provide information on the interplay between nucleosynthesis and mixing processes at different stellar burning sites \cite{harris1988, dearborn1992, nollett2003, abia2011, palmerini2011}.

The two reactions play a similarly important role in the hot-CNO cycles during explosive hydrogen burning in classical novae. At these conditions hydrogen burning reaches peak temperatures around T$=0.1-0.4$ GK \cite{novae}. In this scenario, $^{17}$O$(p,\gamma)^{18}$F directly affects the production of the $\beta^+$~unstable $^{18}$F ($t_{1/2}=110$ min) \cite{iliadis2002, fox2004}, whose $511$ keV electron-positron annihilation $\gamma$-ray could potentially be detected by $\gamma$-ray satellites, such as the INTEGRAL observatory \cite{integral}.

The stellar reaction rates are determined by the reaction cross section at the stellar energy range. At low energies, the reaction rate of $^{17}$O$(p,\gamma)^{18}$F ($Q=5.606$ MeV) is affected by two low lying narrow resonances ($\Gamma<1$ keV) at $E_{r}=65.1$ keV and $E_{r}=183$ keV ($E_{p}=68.9$ and $193$ keV respectively), which have been subject to a number of recent low-energy studies. The resonance at $E_{p}=68.9$ keV is extremely hard to measure directly with current techniques, and its strength is estimated indirectly by experimental and theoretical constrains on its partial widths. On the other hand, recent work by Fox \textit{et al.}~\cite{fox2005} and Chafa \textit{et al.}~\cite{chafa2007} have successfully measured directly the strength of the resonance at $E_{p}=193$ keV. 

This paper reports on a new measurement of the $^{17}$O(p,$\gamma$)$^{18}$F reaction. The measurement focused on examining the nature and strength of cross section components which extend into the low-energy range and contribute substantially to the stellar rate. These contributions include a strong direct capture component and the low-energy tails of two broad resonances at $E_{r}=557$ keV and $E_{r}=677$ keV ($E_{p}=590$ and $717$ keV, respectively). These components had first been investigated by Rolfs \cite{rolfs1973}. The more recent studies by Fox \textit{et al.}~\cite{fox2005} and Chafa \textit{et al.}~\cite{chafa2007} contradict the earlier cross section data and differ by up to a factor of 2. Measurements by Newton \textit{et al.}~\cite{newton2010} and Hager \textit{et al.}~\cite{hager2012} focused on the direct capture component at low energies, in the range $E_{c.m.}=250-500$ keV. The limited energy range in these studies prohibits a direct normalization to the strength of the higher energy resonances. Newton \textit{et al.} estimated the contribution of the two broader resonances in their measured energy range by multiplying the previously recorded resonance strengths \cite{rolfs1973} by a factor of $0.62$, following an earlier suggestion by Fox \textit{et al.} \cite{fox2005}.

The inconsistencies between the different data sets impacts the extrapolation of the $S$-factor to the stellar energy range, which makes a further re-investigation of the reaction necessary. The present study aims at constraining the low-energy contributions of the high energy resonances and the direct capture by measuring the capture cross section of individual primary transitions over a wide range of energies, $E_{p}=365-1800$ keV ($E_{c.m.}=345-1700$ keV) and angles, $\theta_{lab}=0^\circ$, $45^\circ$, $90^\circ$, and $135^\circ$. The data are then fitted within the framework of a multi-level, multi-channel $R$-matrix approach~\cite{azure}, allowing for a more consistent and better-constrained extrapolation to lower energies. The angular distribution information is important in constraining the direct capture contributions of different final state orbital angular momenta, when more than one are possible, as well as the magnitude of the background poles, when these were included. Target yield deconvolution and angular attenuation effects are carefully taken into account and coincidence summing corrections are kept to a minimum, in order to obtain more reliable data. Strengths and branching ratios are provided for several resonances in this energy region, and are compared to literature values.

Sec.~\ref{SEC:experimental} and \ref{SEC:results} describe the experimental equipment, procedure and results. In Sec.~\ref{SEC:rmatrix}, we discuss the $R$-matrix fit of the data and the extrapolation towards lower energies in comparison  with previous studies. The new reaction rate is presented in Sec.~\ref{SEC:rates}. Throughout the paper, $E_{c.m.}$ refers to the reaction energy at the center-of-mass, $E_{p}$ refers to the proton beam energy in the laboratory reference frame, $E_x$ is the excitation energy of the nucleus and $E_{r}$ is the resonance energy in the center-of-mass, unless otherwise noted.

\section{EXPERIMENTAL SETUP AND PROCEDURE \label{SEC:experimental}}

\begin{figure}[t]
\centering
\includegraphics[width=0.48 \textwidth]{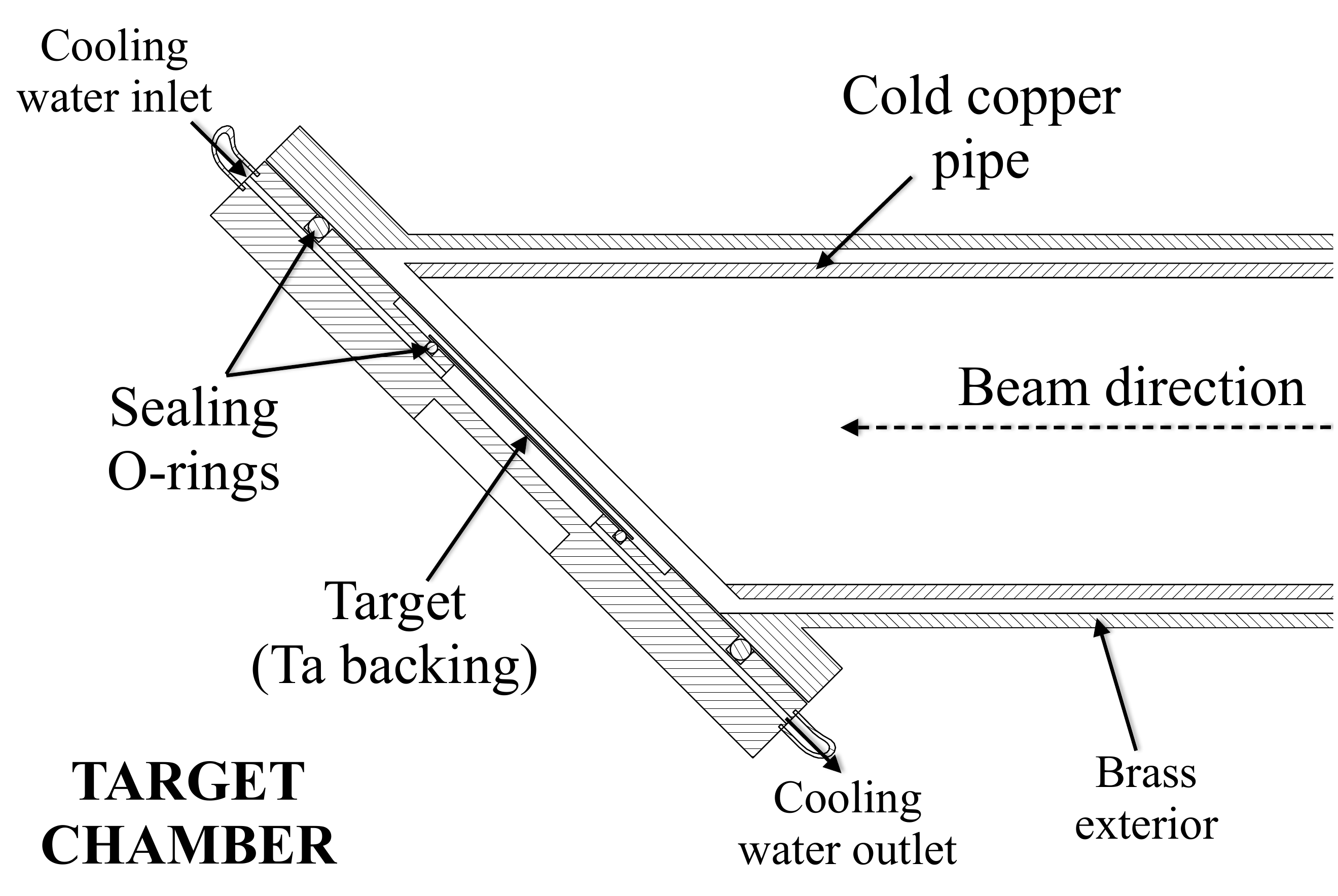}
\caption{Mechanical drawing of the target chamber used for the experiments.}
\label{FIG:chamber}
\end{figure}

The experiments were carried out at the Nuclear Science Laboratory, at the University of Notre Dame. Proton beams were provided by the 4 MV KN Van de Graaff accelerator covering the energy range $E_{p}=600-1800$ keV. For the study of the lower energy range, $E_{p}=365-700$ keV the 1 MV JN Van de Graaff accelerator was used. Typical beam currents between $20$ and $40~\mu$A were achieved for both machines, with a beam energy resolution of approximately $1.5$ keV and an energy uncertainty of $0.5$ keV.
The proton beam entered the target chamber through a liquid-nitrogen-cooled copper pipe (i.e. cold trap) in order to reduce
carbon build-up on the target surface (Fig.~\ref{FIG:chamber}). The cold trap was biased to $-400$ V to suppress secondary electrons from the target. The target and chamber formed a Faraday cup for charge integration. The position of the beam on the target was defined by a set of vertical and horizontal slits. The beam was swept horizontally and vertically across a target area of 2 cm$^2$ by steerers to dissipate power over a large target area. The target was directly water cooled using deionized water.

\begin{figure}[t]
\centering
\includegraphics[width=0.48 \textwidth]{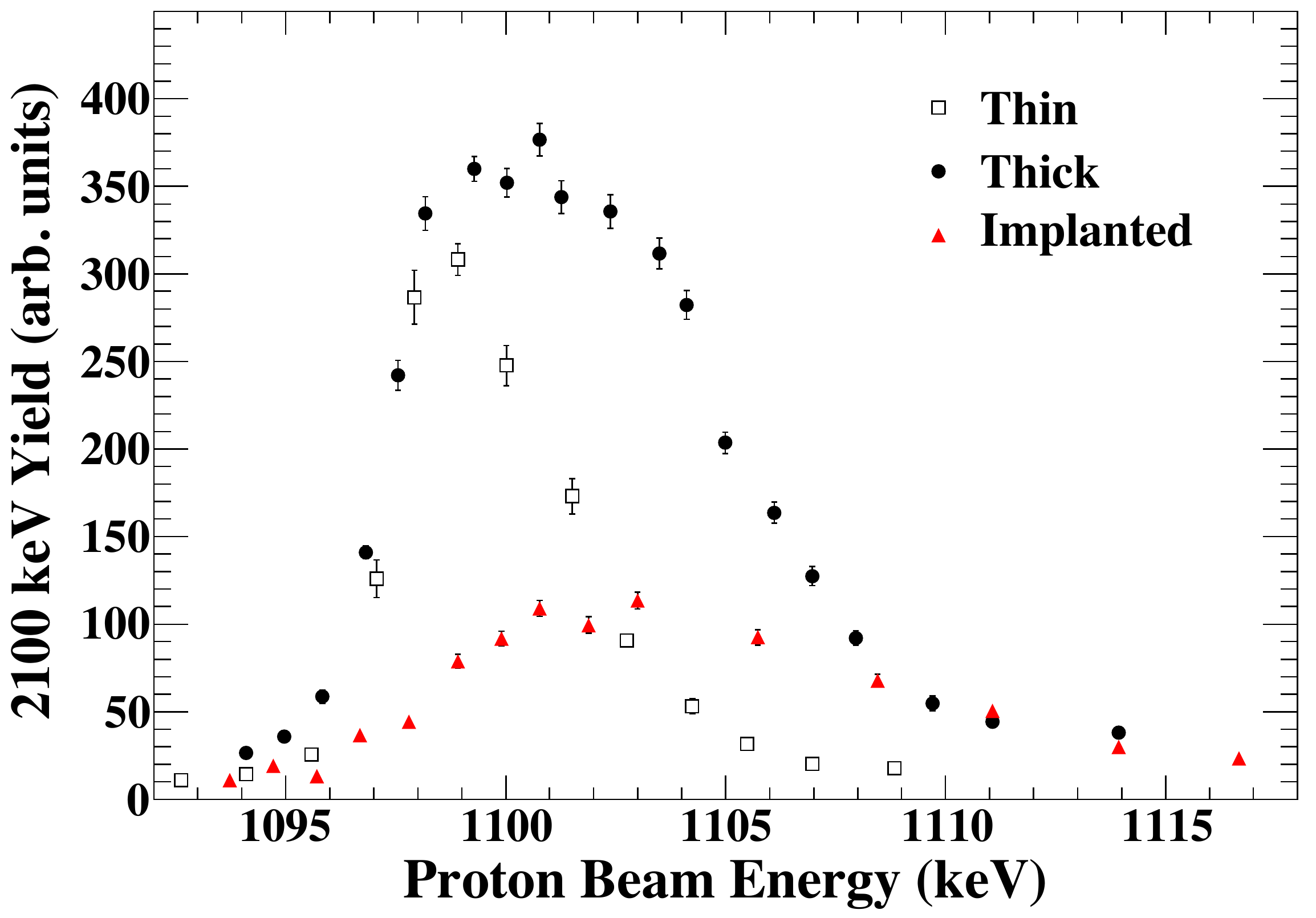}
\caption{(Color online) Yield as a function of energy of the $2100$ keV $\gamma$-ray from the $E_{p}=1098$ keV ($E_{r}=1037$ keV) $^{17}$O$(p,\gamma)^{18}$F narrow resonance. The three sets of data points correspond to the three different targets used in the experiments.}
\label{FIG:scans}
\end{figure}

Two $^{17}$O targets were prepared by anodization of $0.5$ mm sheets of tantalum in $90.1\%$ $^{17}$O-enriched water\footnote{According to the supplier $9.5\%~^{16}$O, $90.1\%~^{17}$O, and $0.4\%~^{18}$O.}, targets Ta$_2$O$_5$ ``Thin'' and ``Thick''. The former was used in the proton energy range $E_{p}=500-1150$ keV, and the latter between $E_{p}=365-1625$ keV. Such targets have been shown~\cite{anodization} to have a well-known stoichiometry (Ta$_2$O$_5$) and their thickness can be controlled by the anodizing voltage. In addition, an implanted target was prepared by bombarding a Ta backing with a $30$ keV $^{17}$O beam. The implanted target was used in the proton energy range $E_{p}=400-180$ keV. Several scans of narrow resonances were performed in order to measure the thickness of all the targets (Fig.~\ref{FIG:scans}). The thickness of target Ta$_2$O$_5$ ``Thick'', prepared with the higher anodization voltage, was determined by the width of the $\gamma$-ray yield curves taken for the narrow resonances at $E_{p}=519$~keV ($\Delta E_{p}=12.5\pm0.8$ keV) and $E_{p}=1098$ keV ($\Delta E_{p}=8.45\pm0.55$~keV). Using stopping power tables~\cite{srim} and the known stoichiometry, the target thickness of Ta$_2$O$_5$ ``Thick'' was measured to be $6.38\pm0.37\times10^{17}$~atoms/cm$^2$. The uncertainty of the thick target was calculated from the uncertainty of the measured energy loss ($\sim5\%$) and the tabulated stopping powers ($\sim4\%$). The uncertainty in the $^{17}$O water enrichment was also taken into account ($\sim3\%$). The thicknesses of the other anodized target, Ta$_2$O$_5$ ``Thin'', and the implanted target were calculated relative to the thickness of Ta$_2$O$_5$ ``Thick'', with values of $3.00\pm0.32\times10^{17}$~atoms/cm$^2$ and $2.25\pm0.26\times10^{17}$~atoms/cm$^2$, respectively. The uncertainties for these targets arise mainly from the uncertainty of the thick target ($\sim6\%$) and the uncertainties of the measured integrated yields of the resonance scans ($\sim8\%$). Resonance scans were repeated periodically throughout the experiments to monitor the quality of the targets. No target degradation was observed. Due to the unknown stoichiometry of the implanted target, only data from the high energy region ($E_{p}>1250$ keV) where the cross section varies smoothly with beam energy were considered in the analysis. The use of the two anodized targets with different thicknesses helped to confirm the yield deconvolution method, as discussed later.

At this point, it is important to note some of the limiting factors in the detection of $\gamma$-rays in the present experiment. First, the decay scheme of the compound nucleus, $^{18}$F, is rather fragmented, which means that an appreciable percentage of decay paths consist of multiple transitions with low branching ratios. As a result, some transitions may have low statistics and may be undetectable because of the large Compton continuum from the stronger transitions. The latter is even more pronounced when using crystals of small volume, as in this work. Therefore, obtaining a full excitation curve of the two broader resonances, $E_{p}=590$ and $717$ keV, including their tails, for all the branchings was not possible in the timeframe of the experiments, and given the background involved. Instead, long runs on-resonance were used to obtain the branching ratio information, as discussed in the next section.

At $\gamma$-ray energies $E_\gamma=5.0-6.2$ MeV, the spectrum is dominated by the presence of the $6130$ keV line from the $^{19}$F$(p,\alpha\gamma)^{16}$O reaction, whose cross section exhibits several strong resonances. This is a common problem when using tantalum backings, since fluorine is part of the extraction process of tantalum, making it very hard to obtain tantalum with sufficiently low fluorine contamination levels. Heating of the tantalum backing in a high vacuum chamber did not decrease the $^{19}$F contamination levels noticeably. For this reason, the detection of the primaries R/DC$\rightarrow 937$ and R/DC$\rightarrow 1121$ was inhibited for a wide proton energy range but more importantly at energies $E_{p}=1250-1450$ keV ($E_{c.m.}=1180-1370$ keV), where the $^{19}$F$(p,\alpha\gamma)^{16}$O reaction exhibits strong resonances. $^{18}$O contamination did not affect the measurements significantly, since the enriched water used for the production of the anodized targets contained only a very small percentage of $^{18}$O. Cross section plots are presented in Sec.~\ref{SEC:rmatrix} and in the appendix along with the $R$-matrix fits.

\begin{figure}[b]
\centering
\includegraphics[width=0.48 \textwidth]{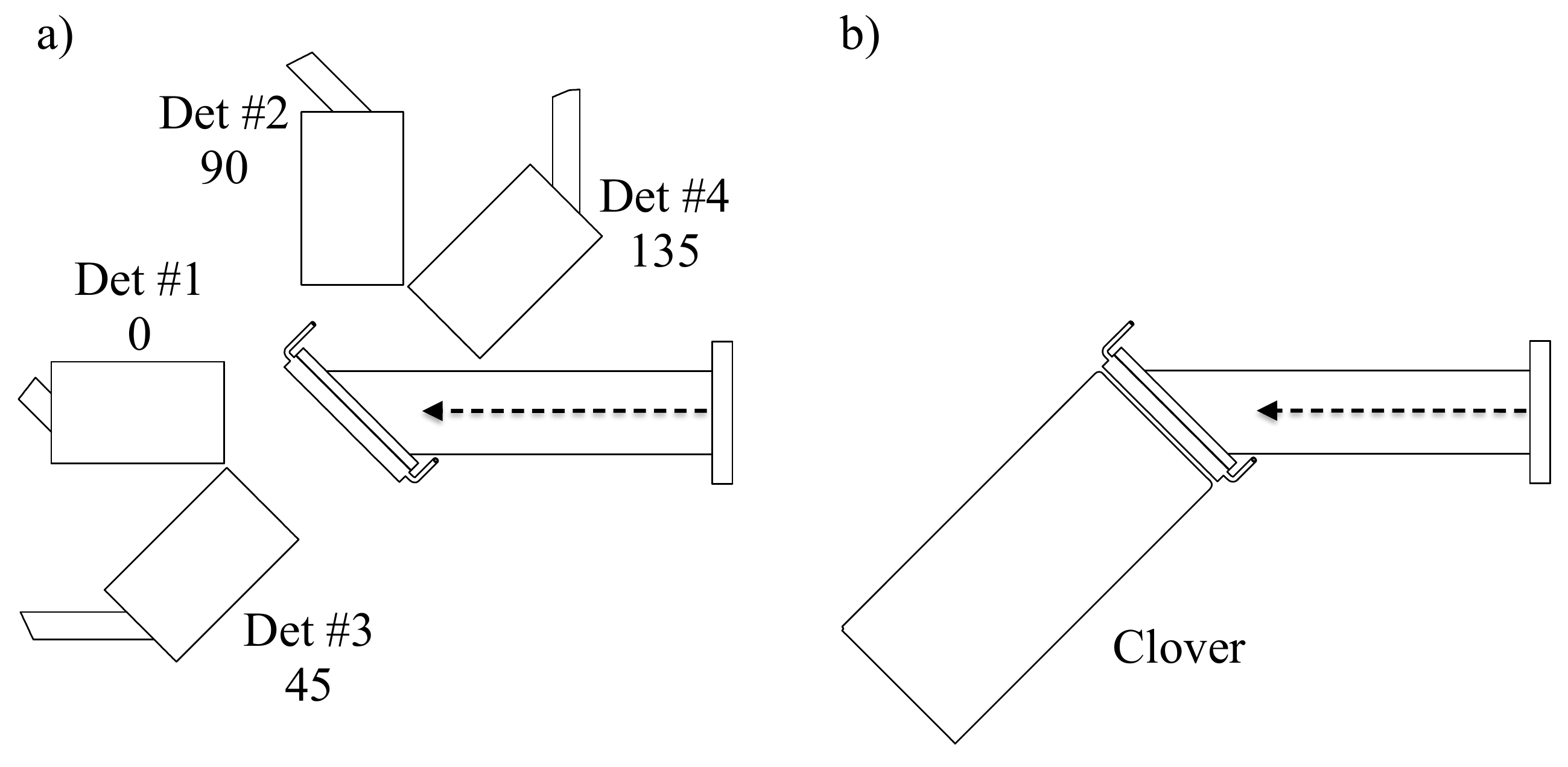}
\caption{Detector geometries of a) the angular distribution setup and b) the clover setup.}
\label{FIG:detectors}
\end{figure}
\begin{figure*}[t]
\centering
\includegraphics[width=0.99 \textwidth]{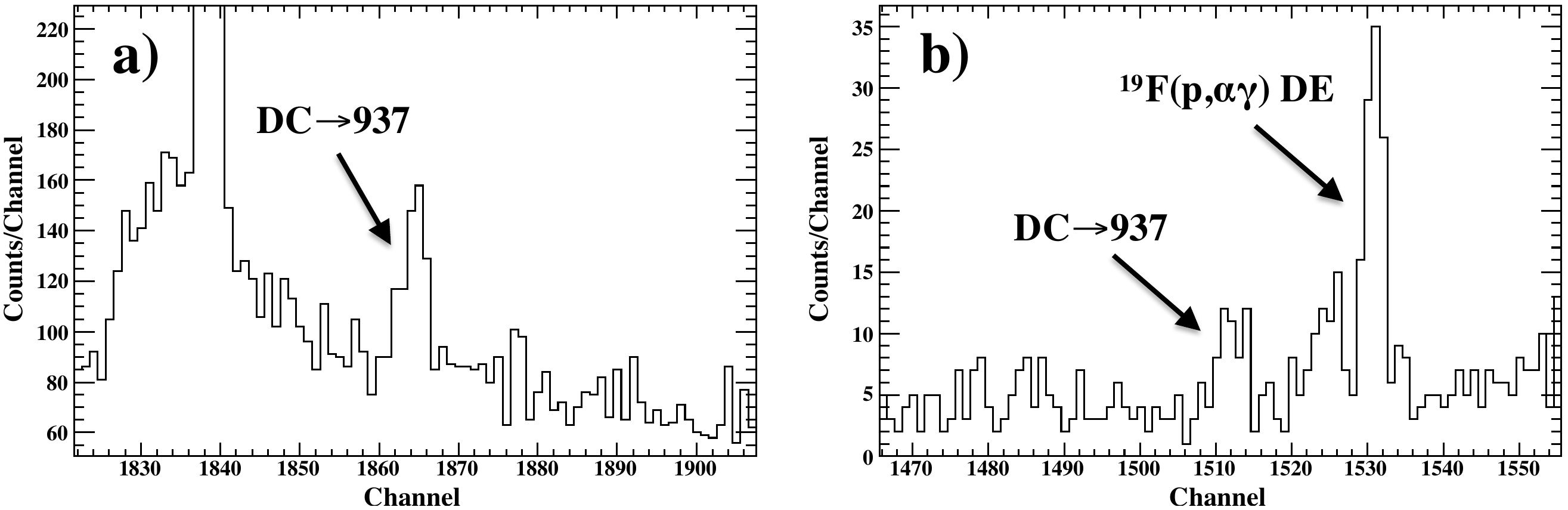}
\caption{Spectra of the $45^\circ$ detector of the angular distribution setup in the energy region of the primary transition to the first excited state in $^{18}$F ($E_x=937$ keV) for proton beam energies of a) $1625$ keV ($E_{c.m.}=1535$ keV) and b) $400$ keV ($E_{c.m.}=378$ keV). The low counting statistics of the low-energy spectrum is indicative of the low cross sections, in combination with the detector geometry. The more prominent peaks in both spectra are due to the $^{19}$F$(p,\alpha\gamma)^{16}$O reaction, from the fluorine contamination in the tantalum backing.}
\label{FIG:spectra}
\end{figure*}
Angular distributions of the prompt $\gamma$-rays from the $^{17}$O$(p,\gamma)^{18}$F reaction were measured with four $20\%$ (relative to a $3^{\text{"}}\times3^{\text{"}}$ NaI detector) high-purity Ge (HPGe) detectors placed at $0^\circ$, $45^\circ$, $90^\circ$, $135^\circ$ at $8$ cm distance from the center of the target (Fig.~\ref{FIG:detectors}a). Two typical spectra showing the primary transition to the first excited state in $^{18}$F ($E_x=937$ keV) obained with the $45^\circ$ detector are shown in Fig.~\ref{FIG:spectra}. The spectrum in Fig.~\ref{FIG:spectra}a was recorded at proton energy $E_{p}=1625$ keV ($E_{c.m.}=1535$ keV), where the cross section is dominated by the direct capture mechanism with an approximate value of $\sim200$ nb/sr. The spectrum in Fig.~\ref{FIG:spectra}b was recorded at proton energy $E_{p}=400$ keV ($E_{c.m.}=378$ keV), which corresponds to the low-energy side of the $E_{p}=590$ keV resonance with an approximate cross section of $\sim4$ nb/sr. In addition, measurements were performed with a HPGe clover detector, segmented in four $25\%$ crystals, placed at $45^\circ$ and $1$ cm from the target (Fig.~\ref{FIG:detectors}b). The spectra of the four individual crystals were added off-line and were analyzed as one. Performing the experiment with two different detector geometries was important in testing the validity of the data analysis procedure. The intrinsic energy resolution for all crystals is $2.2$ keV at $E_\gamma = 1.33$ MeV.

Full peak detection efficiencies as functions of the $\gamma$-ray energy were obtained for both setups using calibrated ($^{60}$Co, $^{137}$Cs) and uncalibrated ($^{56}$Co) $\gamma$-ray sources as well as the well-known narrow resonances of $^{27}$Al$(p,\gamma)^{28}$Si and $^{14}$N$(p,\gamma)^{15}$O at $E_{p}=992$ keV ($E_{c.m.}=957$ keV)~\cite{aluminum} and $278$ keV ($E_{c.m.}=260$ keV)~\cite{imbriani2005}, respectively. Detection efficiencies for the single and double escape peaks were also measured, since most primaries of interest exhibited strong escape lines due to the high Q-value of the reaction.
\begin{figure}[b]
  \begin{center}
    \centerline{\includegraphics[width=0.47 \textwidth]{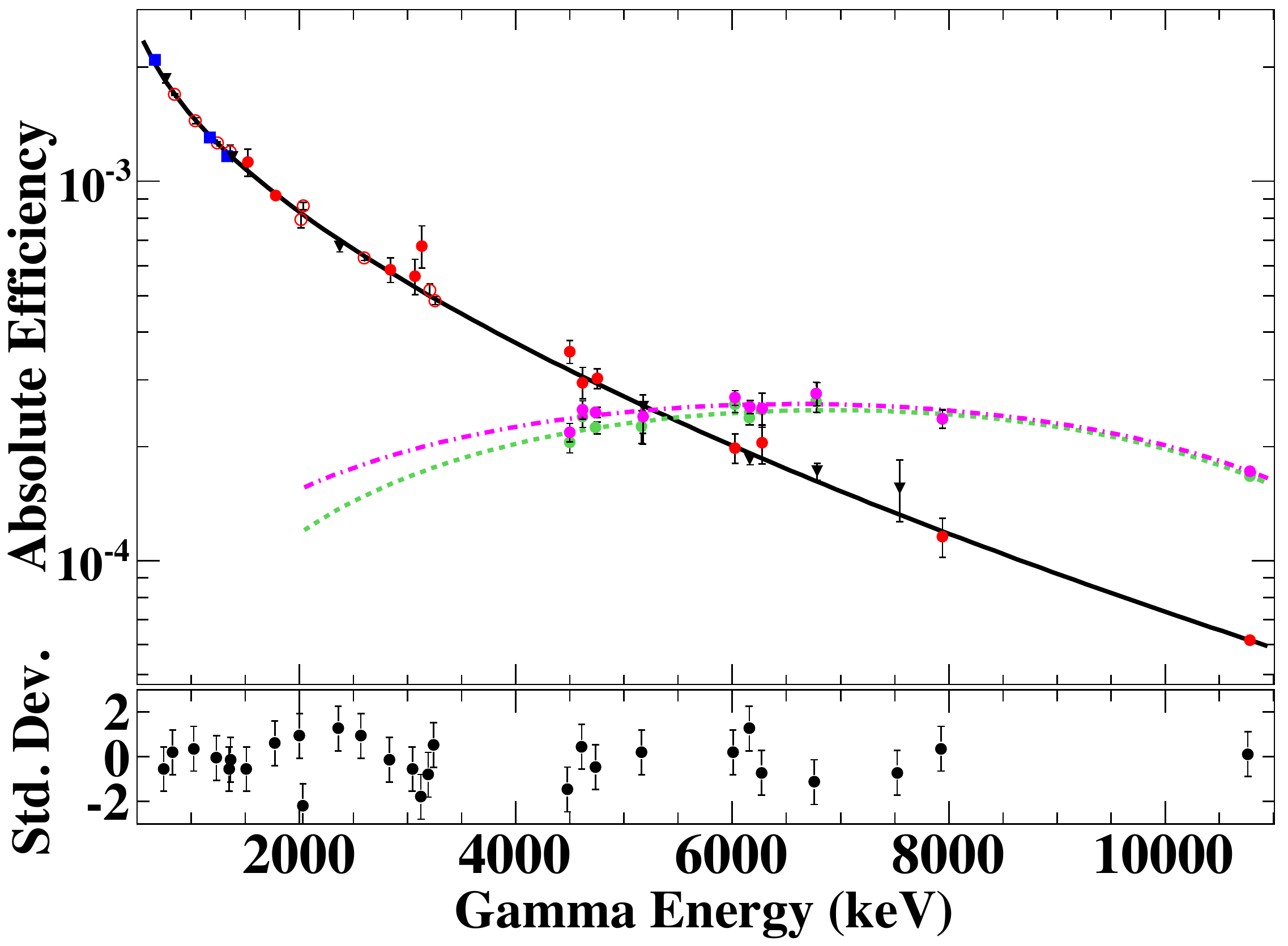}}
    \caption{(Color online) Peak efficiency curves for the $0^{\circ}$ detector from the angular distribution experiment. The red circles refer to $\gamma$ strength measurements of transitions in the $^{27}$Al$(p,\gamma)^{28}$Si reaction, the blue squares refer to $\gamma$ decay measurements with $^{60}$Co  and $^{137}$Cs sources, the black triangles relate to $\gamma$ strength measurements associated with the $^{14}$N$(p,\gamma)^{15}$O reaction and the open red circles refer to $^{56}$Co source measurements. The two dashed lines correspond to the single (purple dot-dashed) and double (green dashed) escape peak efficiencies. The bottom graph shows the deviation of the full peak efficiency fit from the data points in units of standard deviation.}
    \label{FIG:peakeff}
  \end{center}
\end{figure}
Figure~\ref{FIG:peakeff} shows an example of a full energy peak, a single escape peak and a double escape peak efficiency curve. The peak efficiency curve was fitted with the function~\cite{efficiency1998}
\begin{eqnarray}
	\eta_{pe} = \exp(a + b~\ln(E_\gamma) + c~\ln(E_\gamma)^2 + d~\ln(E_\gamma)^3) ,  \label{EQ:pefunction}
\end{eqnarray}
whereas for the single and double escape curves simple second order polynomials were used. Coincidence summing-in corrections were negligible, given the low peak efficiencies (see Fig.~\ref{FIG:peakeff}). A clear indication of this was the fact that no peak appeared in the aluminum resonance runs in the $\gamma$-ray spectra of the clover detector setup at the $\gamma$-energy of $12541$ keV, where a summing-in peak usually appears from the sum of the strong $1779$ and $10761$ keV transitions in $^{28}$Si. 

Coincidence summing-out corrections for each primary transition were calculated using a Monte Carlo program that required input for the total efficiency of the detector set-up as a function of energy and the level scheme of $^{18}$F (taken from~\cite{sens1973,rolfs1973b}). The program would simulate a million decays and subtract from the $\gamma$-ray of interest according to the coincidence $\gamma$-rays and the total efficiency information. The total efficiency was measured at two energy points with the $^{60}$Co  and $^{137}$Cs sources and was extrapolated to the entire energy range of interest using the Monte Carlo simulation package {\sc Geant4}~\cite{geant4}, normalized to the two data points. Typical total efficiency values for the angular distribution setup were approximately $1.5$\% and for the clover setup $5$\%. Due to the low total efficiencies, summing-out corrections were small, and their contribution to the cross section uncertainty was $<1\%$. 

Angular attenuation coefficients were calculated from the formula~\cite{rose1953}
\begin{eqnarray}
	Q_n=\frac{\int_{0}^{\beta_{max}}P_n(\cos\beta)\eta(\beta,E)\sin\beta~d\beta}{\int_{0}^{\beta_{max}}\eta(\beta,E)\sin\beta~d\beta}, \label{EQ:attenuationcoeff}
\end{eqnarray}
with $\beta$ being the angle relative to the detector's symmetry axis and $\eta(\beta, E)$ the peak efficiency as a function of angle and energy \cite{rose1953}. $Q_n$ does not depend strongly on the absolute scale of the peak efficiency and as a consequence on the energy of the $\gamma$-ray. For the present detector setups, $\eta(\beta, E)$ was estimated from {\sc Geant4} simulations. The results were $Q_1=0.988$, $Q_2=0.963$, $Q_3=0.928$ and $Q_4=0.880$ for the angular distribution setup and $Q_1=0.755$, $Q_2=0.388$, $Q_3=0.067$ and $Q_4=-0.103$ for the clover setup.

\section{ANALYSIS AND RESULTS \label{SEC:results}}

All detectable primary transitions from the $^{17}$O$(p,\gamma)^{18}$F reaction and the secondary transition from the first excited state of $^{18}$F at $937$ keV to the ground state were measured in the proton energy range from $365$ to $1800$ keV. The data from the primaries were used in the $R$-matrix analysis. Secondary transitions were not included in the $R$-matrix analysis but were important for the direct comparison to the Rolfs data. The experimental differential yield for any $\gamma$-ray transition was calculated with the equation
\begin{eqnarray}
	Y_{E_\gamma}(E_{p}, \theta_{lab}) = \frac{N_{E_\gamma}}{4\pi N_{proj}\eta_{pe}(E_\gamma)},  \label{EQ:yield}
\end{eqnarray}
where $E_{p}$ is the energy of the proton beam in the laboratory system, $\theta_{lab}$ is the angle of the detector, $N_{E_\gamma}$ is the number of counts of the $\gamma$-ray peak and $N_{proj}$ are the number of protons incident on target. The last quantity is calculated from the total charge collected on the target throughout each run. The deconvolution of target effects from the experimental yield curves was performed by directly performing an $R$-matrix fit on the experimental yields rather than the cross sections, taking into account the thickness of the targets and the effective stopping power. As a first step, each experimental yield curve is fitted separately, using reasonable $R$-matrix parameter input. The experimental cross section is then estimated for each measured energy point from the fitted yield curve and the corresponding calculated cross section by
\begin{eqnarray}
	\sigma_{exp}(E_i)=\left[ \frac{\sigma_{calc}(E_i)}{Y_{calc}(E_i)}\right]_{FIT} Y_{exp}(E_i).
\end{eqnarray}
In this equation, $Y_{exp}(E_i)$ is the experimental yield given in Eq.~\ref{EQ:yield}, $Y_{calc}(E_i)$ is the yield obtained by the $R$-matrix fit taking into account the target thickness, and $\sigma_{calc}(E_i)$ is the cross section that corresponds to the calculated yield when the target effects are removed. This method cannot be applied for narrow resonances, as these have features that cannot be perfectly reproduced by the code, such as beam energy resolution and straggling. For this reason, the properties of the narrow resonances were studied separately.
\begin{figure}[t]
  \begin{center}
    \centerline{\includegraphics[width=0.49 \textwidth]{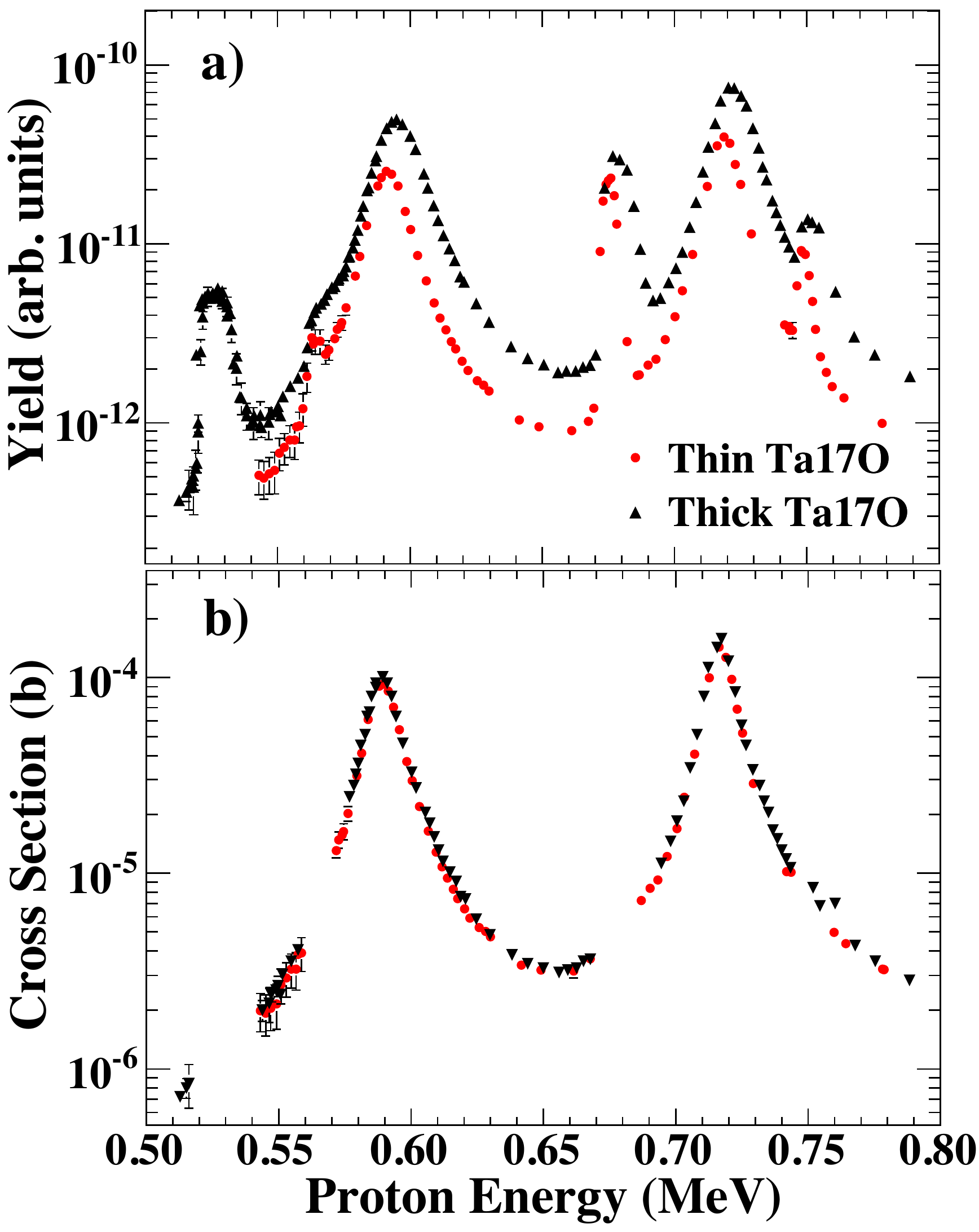}}
    \caption{(Color online) a) Measured yield of the secondary transition from the 1$^{\text{st}}$ excited state to the ground state of $^{18}$F (937$\rightarrow$G.S.) in the energy region of the two broad resonances. The two data sets correspond to the two anodized targets, Ta$_2$O$_5$ ``Thin'' (red circles) and ``Thick'' (black triangles). b)  Converted yield to cross section with the method described in the text. Narrow resonances were removed in this analysis.}
    \label{FIG:937converted}
  \end{center}
\end{figure}

Evidence for the validity of the deconvolution method is presented in Fig.~\ref{FIG:937converted}. Fig.~\ref{FIG:937converted}a shows the measured yield of the secondary transition from the 1st excited state to the ground state of $^{18}$F (937$\rightarrow$G.S.). The black triangles and red circles correspond to the ``Thick'' and ``Thin'' target yield excitation functions, respectively. Fig.~\ref{FIG:937converted}b shows the cross section calculated from the yield with the method described above. The convergence of the two curves is evidence that the procedure works well. A small systematic deviation can be explained by the uncertainties of the target thicknesses. The yields of all observed primary transitions were converted to cross section with this method, excluding most narrow resonances.

 \begin{table*}
 \caption{Resonance strengths in $^{17}$O$(p,\gamma)^{18}$F \label{T:strengths}}
 \begin{ruledtabular}
 \begin{tabular}{ccccccc}
      \multicolumn{1}{c}{$E_{p}$ ($E_{r}$)} & & Present & \multicolumn{4}{c}{Literature} \\
      \multicolumn{1}{c}{(keV)} & J$^\pi$ & $\omega\gamma$ (eV) & Rolfs~\cite{rolfs1973} & Rolfs, scaled\footnote[1]{The resonance strengths given by Rolfs have been corrected by a factor of $0.62$ as suggested by Fox \textit{et al.} in~\cite{fox2005}.} & Sens \textit{et al.} \cite{sens1973} & Fox \textit{et al.} \cite{fox2005} \\
\hline
      519 (490) & $4^-$ & $0.0130\pm0.0015$ & $0.021\pm0.004$ & $0.0130\pm0.0025$    & $0.0068\pm0.0020$ & $0.0137\pm0.0022$ \\
      590 (557) & $3^+$ & $0.37\pm0.05$ & $0.56\pm0.15$ & $0.35\pm0.09$   & $0.18\pm0.05$ & $-$ \\
      717 (677) & $2^+$ & $0.58\pm0.07$ & $0.76\pm0.19$ & $0.47\pm0.12$ &  $0.206\pm0.065$ & $-$ \\
      826 (780) & $2^+$ & $0.0323\pm0.0025$ & $0.050\pm0.015$ & $0.0309\pm0.0093$ &  $0.0175\pm0.0054$ & $-$ \\
      930 (878) & $3^+$ & $0.0194\pm0.0018$ & $0.030\pm0.012$ & $0.0186\pm0.0074$ & $0.010\pm0.005$      & $-$ \\
      1098 (1037) & $2^-$ & $0.297\pm0.033$   & $0.36\pm0.10$ & $0.22\pm0.07$          & $0.130\pm0.043$       & $-$ \\
 \end{tabular}
 \end{ruledtabular}
 \end{table*}

\begin{table*}
\caption{Branchings of certain resonances in $^{18}$F \label{T:branchings}}
\begin{ruledtabular}
\begin{tabular}{ccccccc}
						&	$E_x=6096$	&	6163		&		6283		&		6385		&		6485		&	6644			\\
			Transition		&	$E_{p}=519$	&	590		&		717		&		826		&		930		&	1098			\\
\hline
$	R\rightarrow	$0		&$		-		$&$		-		$&$	0.38	\pm	0.02	$&$	1.4	\pm	0.4	$&$	15.1	\pm	0.7	$&$		-		$\\
$	R\rightarrow	$937		&$	4.2	\pm	0.5	$&$	50.3	\pm	0.6	$&$	65.73\pm	0.35	$&$	74.3	\pm	4.3	$&$	28.7	\pm	2.0	$&$	6.4	\pm	0.5	$\\
$	R\rightarrow	$1040	&$		-		$&$		-		$&$	0.69 \pm 0.07	$&$		-		$&$		-		$&$		-		$\\
$	R\rightarrow	$1121	&$	58.6	\pm	2.3	$&$	2.1	\pm	0.2	$&$		-		$&$		-		$&$	14.2	\pm	1.7	$&$		-		$\\
$	R\rightarrow	$1700	&$		-		$&$		-		$&$	5.67	\pm	0.13	$&$	8.7	\pm	0.5	$&$	7.9	\pm	0.6	$&$		-		$\\
$	R\rightarrow	$2100	&$	25.1	\pm	1.1	$&$		-		$&$	1.74	\pm	0.09	$&$		-		$&$		-		$&$	58.4	\pm	2.7	$\\
$	R\rightarrow	$2523	&$		-		$&$	7.0	\pm	0.2	$&$	0.73	\pm	0.06	$&$		-		$&$	4.7	\pm	0.7	$&$		-		$\\
$	R\rightarrow	$3062	&$		-		$&$	2.3	\pm	0.1	$&$		-		$&$		-		$&$	15.8	\pm	1.6	$&$		-		$\\
$	R\rightarrow	$3133	&$		-		$&$		-		$&$	1.29	\pm	0.06	$&$		-		$&$		-		$&$	22.1	\pm	1.4	$\\
$	R\rightarrow	$3358	&$		-		$&$		-		$&$	2.10 \pm 0.05	$&$		-		$&$		-		$&$		-		$\\
$	R\rightarrow	$3724	&$		-		$&$		-		$&$	2.51	\pm	0.05	$&$		-		$&$		-		$&$	1.6	\pm	0.3	$\\
$	R\rightarrow	$3791	&$	1.3	\pm	0.2	$&$	10.2	\pm	0.2	$&$		-		$&$		-		$&$	4.1	\pm	0.3	$&$	2.8	\pm	0.4	$\\
$	R\rightarrow	$3839	&$		-		$&$	22.0	\pm	0.3	$&$	13.28\pm	0.10	$&$	11.8	\pm	0.8	$&$	8.0	\pm	1.1	$&$		-		$\\
$	R\rightarrow	$4115	&$	1.8	\pm	0.3	$&$	2.2	\pm	0.3	$&$	3.73	\pm	0.06	$&$	3.8	\pm	0.5	$&$		-		$&$	2.2	\pm	0.3	$\\
$	R\rightarrow	$4226	&$		-		$&$	1.4	\pm	0.2	$&$		-		$&$		-		$&$		-		$&$		-		$\\
$	R\rightarrow	$4360	&$		-		$&$		-		$&$	2.14 \pm 0.06	$&$		-		$&$		-		$&$		-		$\\
$	R\rightarrow	$4398	&$	2.2	\pm	0.3	$&$	2.5	\pm	0.1	$&$		-		$&$		-		$&$		-		$&$		-		$\\
$	R\rightarrow	$4652	&$	6.8	\pm	0.3	$&$		-		$&$		-		$&$		-		$&$		-		$&$		-		$\\
$	R\rightarrow	$4860	&$		-		$&$		-		$&$		-		$&$		-		$&$		-		$&$	2.2	\pm	0.3	$\\
$	R\rightarrow	$4964	&$		-		$&$		-		$&$		-		$&$		-		$&$	1.6	\pm	0.2	$&$		-		$\\
$	R\rightarrow	$5502	&$		-		$&$		-		$&$		-		$&$		-		$&$		-		$&$	4.3	\pm	0.3	$\\
\end{tabular}
\end{ruledtabular}
\end{table*}
The strengths of the narrow resonances at $E_{p}=519$, $826$, $930$, and $1098$ keV ($E_{r}=490$, $780$, $878$, and $1037$ keV, respectively) were measured with the Ta$_2$O$_5$ ``Thick'' target, to take advantage of its well-known stoichiometry and thick target yield. Long runs off resonance were also taken in order to identify contributions that do not correspond to the measured resonance and subtracted from the yield data. The angle integrated yield of each transition was obtained by fitting the differential yields of each detector with the equation $W(\theta)=Y_{max}(1+a_2P_2(cos\theta))/4\pi$. The angle integrated yields for each of the observed transitions were added to obtain the branching ratios and the resonance strengths. The latter quantity was calculated using the formula
\begin{eqnarray}
	\omega \gamma_{(p,\gamma)} = \frac{\epsilon_\mathrm{eff,r}~Y_{max}}{\lambda_r^2}\frac{\pi}{\arctan(\Delta E/\Gamma)},
\end{eqnarray}
where $Y_{max}$ is the total angle integrated $\gamma$-ray yield measured on the target yield plateau, $\lambda_r$ is the de Broglie wavelength at the resonance energy, $\Delta E$ is the energy loss of the proton beam inside the target at the resonance energy, $\Gamma$ is the total width of the resonance, taken from Ref.~\cite{tilley1995}, and $\epsilon_\mathrm{eff,r}$ is the effective center-of-mass stopping power of protons in tantalum pentoxide at the resonance energy~\cite{srim}. The strengths of the two broader resonances at $E_{p}=590$~keV and $717$~keV were determined from the $R$-matrix fit parameters, presented in Sec.~\ref{SEC:rmatrix}. Tables~\ref{T:strengths} and \ref{T:branchings} summarize the results for the strengths and branching ratios of the six measured resonances. In Table~\ref{T:strengths} the results are compared with literature data. The values of Rolfs, corrected by a factor of $0.62$ agree very well with the present data. This result confirms the analysis by Fox \textit{et al.}~\cite{fox2005} indicating that the strengths quoted by Rolfs \cite{rolfs1973} have to be renormalized using the updated value for the $E_{p} = 632$ keV resonance strength in the $^{27}$Al$(p,\gamma)^{28}$Si reaction~\cite{iliadis2001}. The measurements by Sens \textit{et al.}~\cite{sens1973} are systematically lower by a factor of 2 compared to the present work, possibly due to the use of unpublished stopping power values. Finally, it is important to note that the resonance strength of the $519$ keV resonance is in excellent agreement with the more recent measurement by Fox \textit{et al.} The experimental uncertainties arise primarily from statistics, the uncertainty in the absolute ($8\%$) and relative ($3\%$) peak efficiency and the uncertainty of the effective stopping power ($\sim4\%$). The branching ratios for the strong transitions reported in Table~\ref{T:branchings} agree very well with values found in literature~\cite{sens1973,rolfs1973b}. For weaker transitions some discrepancies are observed which could point to unaccounted summing effects in the literature values.

The cross section of the secondary transition from the first excited state of $^{18}$F to its ground state ($937\rightarrow$G.S.) was measured to be $\sim40\%$ lower than that of Rolfs (see figure 11 of~\cite{rolfs1973}). Note that this $40\%$ discrepancy is not related to the $0.62$ factor with which the resonance strengths were renormalized, at least to the best of the authors' knowledge. It is not clear if or how much of this discrepancy is related to the erroneous resonance strength value used for the efficiency calibration, since the yield of this transition was measured relative to that of $^{16}$O$(p,\gamma)$. To check the self-consistency of the present analysis, the yield of this transition was also calculated using the measured yield for the primary transitions and the known branching through the $937$ keV $^{18}$F state. The calculated expected yield and the measured yield of the $937$ keV transition were then compared and found to agree to within $5\%$ of each other, a strong indication of a correct efficiency calibration and branching ratio information.

\section{R-MATRIX ANALYSIS AND DISCUSSION \label{SEC:rmatrix}}

A multi-channel, multi-level $R$-matrix analysis, as described earlier~\cite{azure}, was used to simultaneously fit all the measured primary transitions of the $^{17}$O$(p,\gamma)$ reaction, as well as the $^{17}$O$(p,\alpha)$ reaction measured by Kieser \textit{et al.}~\cite{kieser1979}. As explained in Sec.~\ref{SEC:results}, the yield data were converted to cross sections by a target effect deconvolution method. The results from that analysis were used as input for the global $R$-matrix fit, as is discussed in this section. Data from most narrow resonances were excluded from the fits since the required number of target integration steps would have slowed down the calculation time substantially,  without improving the accuracy of the fit and extrapolation. The direct capture was assumed to be $E1$ only, as was suggested by previous studies~\cite{fox2005},\cite{rolfs1973}. The $R$-matrix radius for the proton channel was taken as $r_c=r_{0}\times (A_t^{1/3}+A_p^{1/3}) = 4.46$ fm, with $r_{0}=1.25$ fm. The $R$-matrix radius for the $\alpha$-channel was taken to be $\alpha_c=5.44$ fm. The choice of $r_{0}=1.25$ fm is common in literature (e.g.~\cite{iliadis2001}), and was adopted as a reasonable estimate. The effect of the channel radius value on the results is discussed later.

In $R$-matrix theory the direct capture (DC) part of the cross section is divided into an external capture part and an internal part. The external capture component has a given energy dependence, whereas its magnitude is defined by the \emph{asymptotic normalization coefficient} (ANC) of the bound state. The ANCs can either be treated as parameters of the fit, or they can be derived from experimental spectroscopic factors ($C^2S$) of the bound states taken from literature. The internal part is treated with high energy background poles.

For each of the observed primary transitions the ANC of the final state is one of the $R$-matrix parameters. Different than resonance parameters these parameters are only restricted by the respective cross sections. Because ANC values are only poorly restricted from the present data set, literature values derived from transfer reactions \cite{polsky1969,landre1989} were taken where appropriate. 

To test if these values are consistent with our data, two $R$-matrix fits were performed: a \emph{test} fit and a \emph{main} fit. In the test fit, no background poles were included and the ANCs were left as free parameters. The resulting ANCs were converted to spectroscopic factors using the equation
\begin{eqnarray}
	 C^2S = \left(\frac{C_{fit}}{b_{sp}}\right)^2,  \label{sanc}
\end{eqnarray}
where $C^2S$ is the proton single particle spectroscopic factor of a given bound state of $^{18}$F with the isospin Clebsch-Gordan coefficient factored in, $C_{fit}$ is the ANC value obtained by the fit, and $b_{sp}$ is the ANC value obtained by a simple single particle mean field calculation, using a Wood-Saxon potential with radius $r=1.25 A_T^{1/3}$ fm, diffuseness $a=0.65$ fm and depth chosen to reproduce the binding energy of the state. The spectroscopic factors $C^2S$ were then compared to literature spectroscopic factors from transfer reactions~\cite{polsky1969,landre1989}\footnote{The bound state potential used in the study by~\cite{polsky1969} is identical to the one used in this study to calculate the single particle spectroscopic factor. The potential used by~\cite{landre1989} is similar to this study's with the exception of a spin-orbit term.} to determine their validity (Table~\ref{T:anc}). If a fitted ANC of a given transition corresponds to a spectroscopic factor that is much larger than the one found in literature, that indicates that the transition cannot be fully described by the external capture component of the $R$-matrix theory, and a background pole is required. 

In the main fit, the spectroscopic factors that disagreed with the literature data were fixed close to experimental spectroscopic factor values, and the missing contributions were replaced by high energy background poles. These poles were placed at an excitation energy of $15$ MeV, which is about $8$ MeV above the energy region of the experimental data, with spins and parities chosen so that primary transitions of interest decay via $E1$. As there are no available proton scattering data or other appropriate experimental data to constrain the proton partial widths of the poles, they were fixed at $\Gamma_p=6$ MeV, chosen close to the Wigner limit given by the formula $\Gamma_W=2P_l(E_{r})\gamma_W$, where $P_l(E_{r})$ is the penetrability factor for a given partial wave at the resonance energy and $\gamma_W=(3/2)\hbar^2/(\mu r_c^2)$. The $\Gamma_\gamma$ partial width was left as a fit parameter. 

\begin{figure}[t]
\centering
\includegraphics[width=0.48 \textwidth]{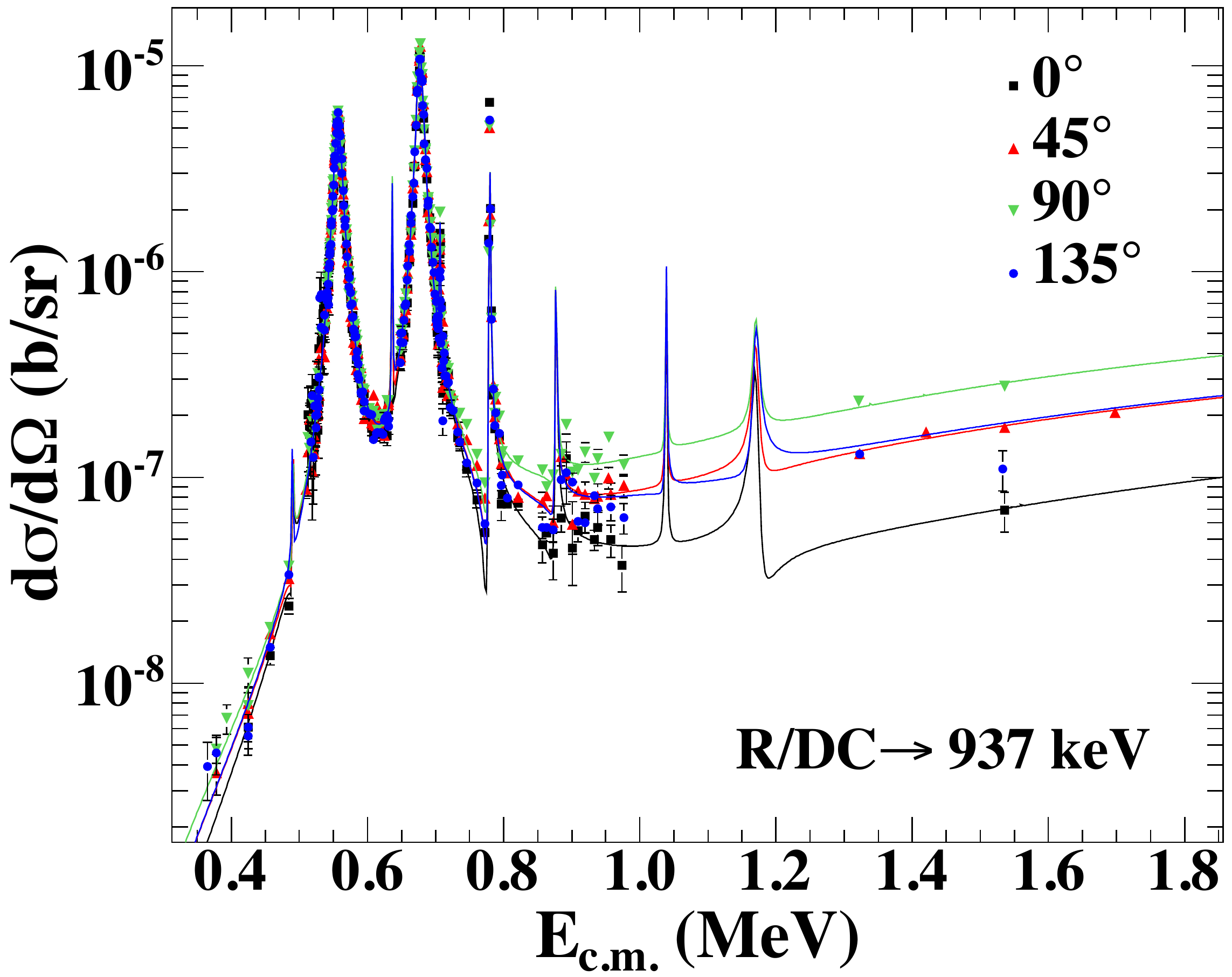}
\caption{(Color online) $R$-matrix fit of the differential cross section of the $^{17}$O$(p,\gamma)^{18}$F*$(937$ keV$)$. The different colors and shapes correspond to different angles.}
\label{FIG:fit937}
\end{figure}
\begin{figure}[t]
\centering
\includegraphics[width=0.48 \textwidth]{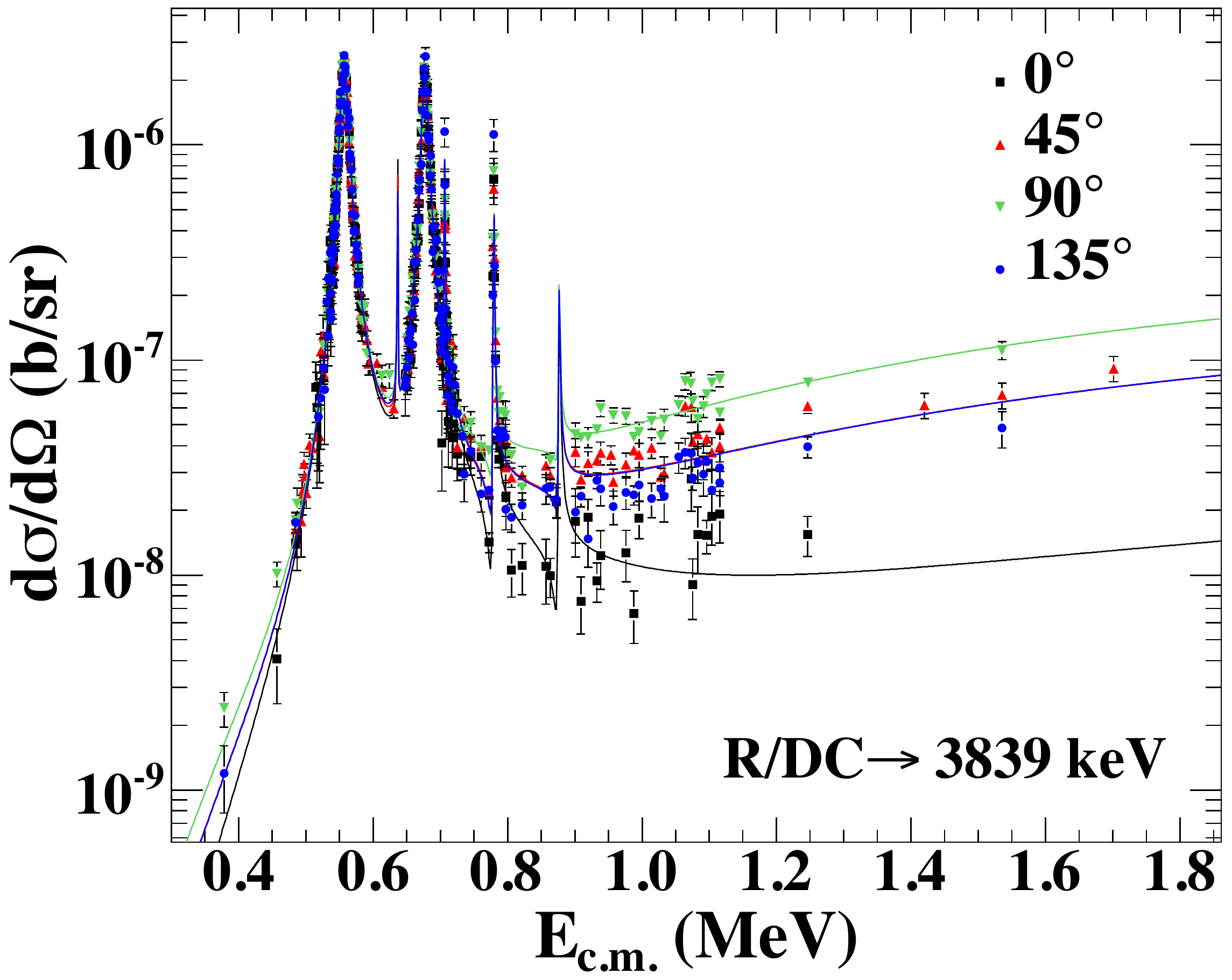}
\caption{(Color online) $R$-matrix fit of the differential cross section of the $^{17}$O$(p,\gamma)^{18}$F*$(3839$ keV$)$. The different colors and shapes correspond to different angles. The $45^\circ$ and $135^\circ$ lines lie on top of each other and cannot be distinguished.}
\label{FIG:fit3839}
\end{figure}
The fits of the two strongest transitions, R/DC$\rightarrow937$ and R/DC$\rightarrow3839$ keV are shown in Figs.~\ref{FIG:fit937} and \ref{FIG:fit3839}.  Their contribution to the low-energy total $S$-factor amounts to ($\sim40\%$) and ($\sim16\%$) respectively. The fits of the weaker transitions are shown in the appendix. The reduced-$\chi^2$ of all the fits ranged between $1.5$ and $2.7$, with the exception of the R/DC$\rightarrow937$ fit that had a value of $6.5$. The high reduced-$\chi^2$ values are due to the fact that this is a simultaneous fit of multiple transitions, multiple angles and two targets of different thicknesses. In addition, imperfect target effect corrections have a big effect on the reduced-$\chi^2$ when regions of sharp changes in the cross section are included, as are in the R/DC$\rightarrow937$ transition.

A good fit was also obtained in this global fit for the $^{17}$O$(p,\alpha)^{14}$N, shown in Fig.~\ref{FIG:kieser} in the appendix. No target effects were taken into account in the case of the $(p,\alpha)$ channel, as that would slow down the calculations. This leads to a poor fit for the narrow resonances, however, it does not affect the results for the two broad resonances at $E_{r}=557$ and $677$ keV. This was verified by a separate fit of the $(p,\alpha)$ data alone with target effects included. This led to a better description of the narrow resonances, but it did not have an influence on the parameters for the broad resonances.

The fitted and fixed ANC values from all the fits and the respective calculated spectroscopic factors are shown in Table~\ref{T:anc}. In bold font are the ANCs that were fixed according to literature spectroscopic factors from~\cite{polsky1969}. The latter values were preferred over those from~\cite{landre1989} since the nuclear potential used in that work was identical to the one used in this work for the conversion of the ANCs to spectroscopic factors. The ANC values in normal font were treated as free parameters, since the corresponding spectroscopic factors were close to the literature values. It is clear that the validity of this approach to estimate the direct capture component depends on the reliability of the ANC values. New measurements to verify the ANC values used in this work would be desirable.
In the present study, an attempt is made to quantify the sensitivity of the extrapolation to the ANC values, as discussed below.

\begin{table}[t]
\caption{ANCs and spectroscopic factors \label{T:anc}}
\begin{ruledtabular}
\begin{tabular}{lcccccccc}
      \multicolumn{2}{c}{Level} 			 & \multicolumn{2}{c}{{\bf Fixed}/Fitted\footnote{Values in bold font were adopted from literature (for details see discussion in text).}} & \multicolumn{2}{c}{Literature $C^2S$}		\\
      \multicolumn{1}{c}{Energy} & $l_f$ &  \multicolumn{1}{c}{ANC ($fm^{-1/2}$)}	& \multicolumn{1}{c}{$C^2S$} 	& \cite{polsky1969} & \cite{landre1989}	\\
      \hline
      937 & $0$    & $\mathbf{6.1}$ & $\mathbf{0.32}$ & $0.30<x<0.34$ & $0.10\footnote[2]{According to~\cite{landre1989} these spectroscopic factors have a large uncertainty due to a broad $\chi^2$ minimum.}$ \\
      937 & $2$    & $\mathbf{1.2}$ & $\mathbf{0.18}$ & $<0.36$ & $0.30\footnotemark[2]$ \\
      1121 & $2$ & $\mathbf{2.7}$ & $\mathbf{1.0}$ & $1.00$ & $0.89$ \\
      2523 & $0$ & $\mathbf{1.4}$ & $\mathbf{0.025}$ & $0.025$ & $0.014$ \\
      2523 & $2$ & $-$ & $-$ & $<0.005$ & $0.011$ \\
      3062 & $0$ & $4.5$ & $0.32$ & $0.16$ & $0.21$ \\
      3062 & $2$ & $1.0$ & $0.37$ & $0.74$ & $0.62$ \\
      3839 & $0$ & $4.6$ & $0.42$ & $0.50$ & $-$ \\
      3839 & $2$ & $0.6$ & $0.19$ & $<0.56$ & $-$ \\
      4115 & $0$ & $\mathbf{2.5}$ & $\mathbf{0.13}$ & $0.13$ & $0.17$ \\
      4115 & $2$ & $\mathbf{1.0}$ & $\mathbf{0.75}$ & $0.75$ & $0.68$ \\
      4652 & $2$ & $1.3$ & $1.6$ & $0.98$ & $1.04$ \\
      4964 & $0$ & $3.2$ & $0.17$ & $0.22$ & $0.17$ \\
      4964 & $2$ & $0.7$ & $0.5$ & $0.49$ & $0.51$ \\
\end{tabular}
\end{ruledtabular}
\end{table}

The resulting resonance parameters of the main fit are shown in Table~\ref{T:parameters}. Values in parentheses denote fixed parameters. The energies of the two broad resonances are in excellent agreement with literature values. The proton width of the $6.16$ MeV state, $\Gamma_p=14.1\pm 0.3$ keV, agrees very well with the reported literature values of $13\pm1$ keV~\cite{rolfs1973b} and $14.7\pm1.5$ keV~\cite{sens1973}, whereas some disagreement is observed for the proton width of the $6.28$ MeV state, $\Gamma_p=11.3\pm 0.2$ keV, compared to the literature values, $8.0\pm2.0$ keV~\cite{rolfs1973b} and $8.5\pm1.0$ keV~\cite{sens1973}. The reported uncertainties include only the statistical uncertainties that arise from the fit. In addition, there are systematic errors arising from the choice of fixed $R$-matrix parameters (e.g. channel radii, ANC). Since these parameters are correlated, these systematic errors are difficult to extract. However, deviations greater than $0.2$ keV were never observed from the various fits. Additional background poles were also tested, in order to improve the fits of the two broader resonances, but no significant effects were observed. 

\begin{table*}
\caption{Parameters obtained from the $R$-matrix fit \label{T:parameters}}
\begin{ruledtabular}
\begin{tabular}{cccccccccccc}
			& $\Gamma_p$ 	& $\Gamma_\alpha$ 	& \multicolumn{8}{c}{$\Gamma_\gamma$ (meV)}	\\
$E_x$ (MeV) 	& (keV)	& (eV) 	&	$	R\rightarrow	$937		& $	\rightarrow	$1121	& $	\rightarrow	$1700	&  $	\rightarrow	$2523	&  $	\rightarrow	$3791	& $	\rightarrow	$3839	& $	\rightarrow	$4115	& $	\rightarrow	$4652	& $	\rightarrow	$4964\\
\hline
$6.16$ $(3^+)$	&$	14.1\pm0.3	$&$	8\pm1	$&$	318\pm4	$&$	-	$&$	-	$&$	47\pm2	$&$	69\pm2	$&$	140\pm3	$&$	20\pm3	$&$	-	$&$	-	$\\
$6.28$ $(2^+)$	&$	11.3\pm0.2	$&$	30\pm4	$&$	930\pm10	$&$	-	$&$	130\pm15	$&$	-	$&$	-	$&$	175\pm5	$&$	-	$&$	-	$&$	-	$\\
$(15)$ $(2^-)$		&$	(6\times10^3)	$&$	-	$&$	-	$&$	-	$&$	-	$&$	8.0\times10^{6}$&$	-	$&$	-	$&$	-	$&$	-	$&$	-	$\\
$(15)$ $(3^-)$		&$	(6\times10^3)	$&$	-	$&$	3.2\times10^{6}$&$	-	$&$	-	$&$	-	$&$	-	$&$	-	$&$	-	$&$	-	$&$	3.4\times10^{6}	$\\
$(15)$ $(4^-)$		&$	(6\times10^3)	$&$	-	$&$	2.6\times10^{6}	$&$	2.1\times10^{6}	$&$	-	$&$	-	$&$	-	$&$	-	$&$	3.0\times10^{6}	$&$	-	$&$	-	$\\
\end{tabular}
\end{ruledtabular}
\end{table*}
 
In order to make the comparison with previous work easier, the extrapolated cross section curve was converted to the astrophysical $S$-factor curve given by the formula
\begin{eqnarray}
	 S(E) = \sigma(E)E e^{2\pi\eta},  \label{sfactor}
\end{eqnarray}
where $\eta=0.1575\times Z_1Z_2\left(\mu/E\right)^{1/2}$ is the Sommerfeld parameter, with $\mu$ in units of u, and $E$ in MeV. 
\begin{figure}[!t]
\centering
\includegraphics[width=0.48 \textwidth]{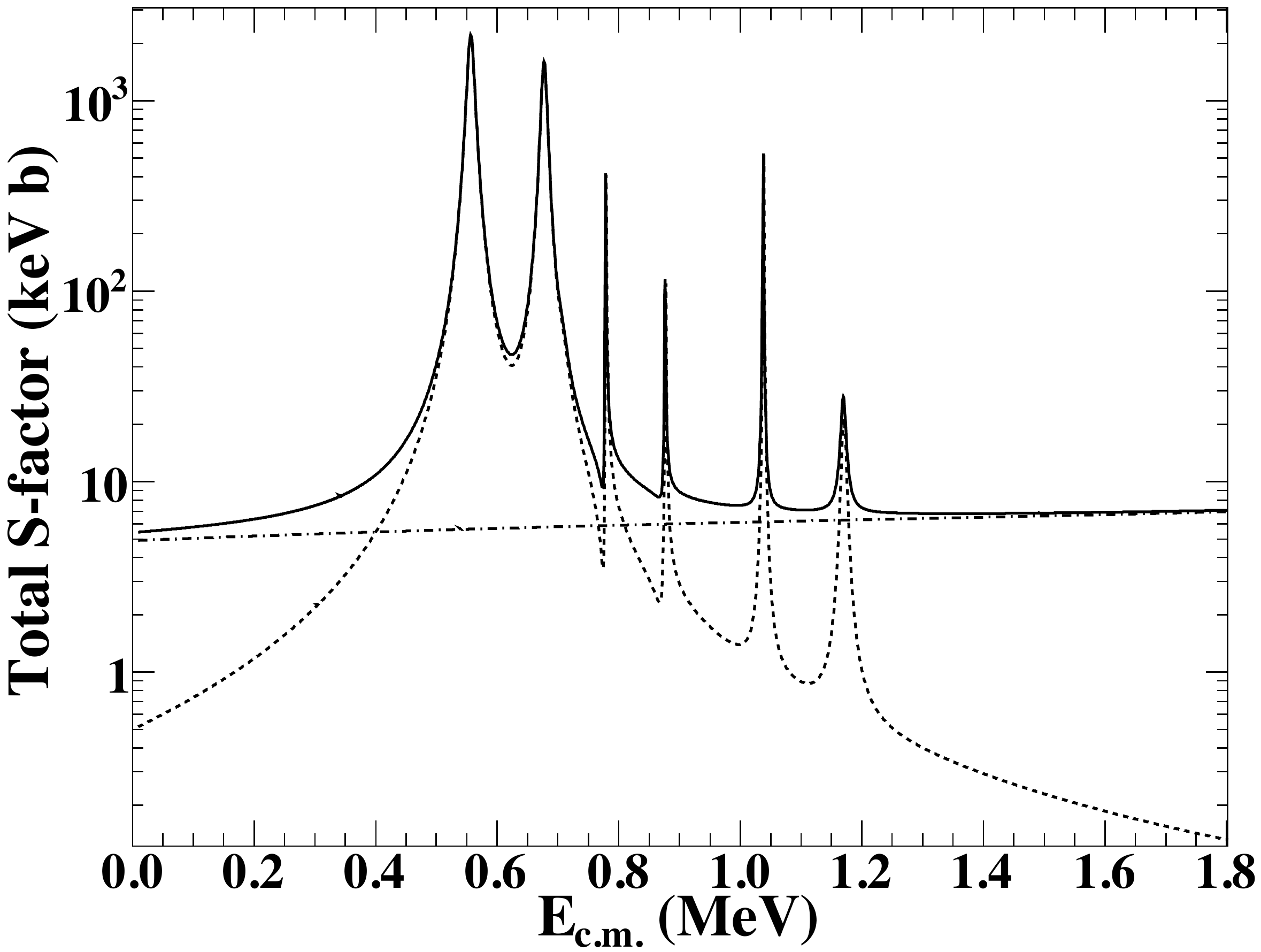}
\caption{$S$-factor obtained from the $R$-matrix fit. The continuous line represents the total $S$-factor from the best $R$-matrix fit, with background poles included and some ANC values fixed. The dashed-dotted line and the dashed line are the contributions of the direct capture (background poles included) and the high energy resonances respectively. Some narrow resonances are omitted from this plot.}
\label{FIG:sfactor}
\end{figure}
\begin{table}[h]
\caption{Calculated S-factors \label{T:sfactor}}
\begin{ruledtabular}
\begin{tabular}{cccc}
      \multicolumn{1}{c}{$E_{c.m.}$} 			& Direct & Resonant & Total	 \\
      \multicolumn{1}{c}{(keV)} 			& \multicolumn{1}{c}{(keV b)} & \multicolumn{1}{c}{(keV b)} & \multicolumn{1}{c}{(keV b)}	 \\
      \hline
$	10	$ & $	4.92	$ & $	0.52	$ & $	5.4	\pm \text{(th.) }	0.9	\pm \text{(exp.) }	0.7	$ \\
$	20	$ & $	4.93	$ & $	0.54	$ & $	5.5	\pm \text{(th.) }	0.9	\pm \text{(exp.) }	0.7	$ \\
$	40	$ & $	4.96	$ & $	0.58	$ & $	5.5	\pm \text{(th.) }	0.9	\pm \text{(exp.) }	0.7	$ \\
$	60	$ & $	4.99	$ & $	0.62	$ & $	5.6	\pm \text{(th.) }	0.9	\pm \text{(exp.) }	0.7	$ \\
$	80	$ & $	5.01	$ & $	0.68	$ & $	5.7	\pm \text{(th.) }	1.0	\pm \text{(exp.) }	0.7	$ \\
$	100	$ & $	5.04	$ & $	0.74	$ & $	5.8	\pm \text{(th.) }	1.0	\pm \text{(exp.) }	0.7	$ \\
$	120	$ & $	5.07	$ & $	0.80	$ & $	5.9	\pm \text{(th.) }	1.0	\pm \text{(exp.) }	0.7	$ \\
$	140	$ & $	5.10	$ & $	0.88	$ & $	6.0	\pm \text{(th.) }	1.0	\pm \text{(exp.) }	0.7	$ \\
$	160	$ & $	5.12	$ & $	0.96	$ & $	6.1	\pm \text{(th.) }	1.0	\pm \text{(exp.) }	0.7	$ \\
$	180	$ & $	5.15	$ & $	1.07	$ & $	6.2	\pm \text{(th.) }	1.0	\pm \text{(exp.) }	0.7	$ \\
$	200	$ & $	5.18	$ & $	1.18	$ & $	6.3	\pm \text{(th.) }	1.0	\pm \text{(exp.) }	0.8	$ \\
$	220	$ & $	5.21	$ & $	1.31	$ & $	6.5	\pm \text{(th.) }	1.0	\pm \text{(exp.) }	0.8	$ \\
$	240	$ & $	5.23	$ & $	1.47	$ & $	6.7	\pm \text{(th.) }	1.0	\pm \text{(exp.) }	0.8	$ \\
$	260	$ & $	5.26	$ & $	1.66	$ & $	6.9	\pm \text{(th.) }	1.0	\pm \text{(exp.) }	0.8	$ \\
$	280	$ & $	5.28	$ & $	1.89	$ & $	7.1	\pm \text{(th.) }	1.0	\pm \text{(exp.) }	0.9	$ \\
$	300	$ & $	5.31	$ & $	2.18	$ & $	7.4	\pm \text{(th.) }	1.0	\pm \text{(exp.) }	0.9	$ \\
$	320	$ & $	5.34	$ & $	2.53	$ & $	7.8	\pm \text{(th.) }	1.0	\pm \text{(exp.) }	0.9	$ \\
$	340	$ & $	5.36	$ & $	2.98	$ & $	8.3	\pm \text{(th.) }	1.0	\pm \text{(exp.) }	1.0	$ \\
$	360	$ & $	5.39	$ & $	3.55	$ & $	8.9	\pm \text{(th.) }	1.0	\pm \text{(exp.) }	1.1	$ \\
$	380	$ & $	5.41	$ & $	4.32	$ & $	9.6	\pm \text{(th.) }	1.0	\pm \text{(exp.) }	1.2	$ \\
$	400	$ & $	5.44	$ & $	5.38	$ & $	10.7	\pm \text{(th.) }	1.0	\pm \text{(exp.) }	1.3	$ \\
$	420	$ & $	5.47	$ & $	7.00	$ & $	12.3	\pm \text{(th.) }	1.0	\pm \text{(exp.) }	1.5	$ \\
$	440	$ & $	5.49	$ & $	9.36	$ & $	14.6	\pm \text{(th.) }	1.0	\pm \text{(exp.) }	1.7	$ \\
$	460	$ & $	5.52	$ & $	13.24	$ & $	18.3	\pm \text{(th.) }	1.0	\pm \text{(exp.) }	2.2	$ \\
$	480	$ & $	5.54	$ & $	20.34	$ & $	25.2	\pm \text{(th.) }	1.1	\pm \text{(exp.) }	3.0	$ \\
$	500	$ & $	5.56	$ & $	35.65	$ & $	40.0	\pm \text{(th.) }	1.1	\pm \text{(exp.) }	4.8	$ \\
\end{tabular}
\end{ruledtabular}
\end{table}

Figure~\ref{FIG:sfactor} shows the total $S$-factor of the reaction obtained from the main $R$-matrix fit (continuous line). The figure also shows the resonant and direct capture contributions separately in dashed and dashed-dotted lines respectively. As pointed out by~\cite{fox2005} and \cite{chafa2007}, the total direct capture contribution dominates at energies less than $E_{c.m.}=400$ keV, when compared with the total contribution of the tails of all the high energy resonances. No significant interference was observed between the two components at the low energy side of the two broad resonances, as the two mechanisms involve different initial orbital angular momenta ($l_i=1,3$ for direct, $l_i=0$ for the resonances, known from angular momentum and parity conservation considerations). The two broad resonances decay mostly via M$1$ transitions as opposed to E$1$ transitions for the direct capture. 

The result of the $R$-matrix extrapolation had little dependence on the choice of the channel radius ($\sim3\%$ for the R/DC$\rightarrow$937 extrapolation for a radius $4<r_c<5$ fm) or the position of the background poles ($\sim5\%$ for the total extrapolation when the pole was placed at $35$ MeV instead of $15$ MeV). 
\begin{figure}[!b]
\centering
\includegraphics[width=0.48 \textwidth]{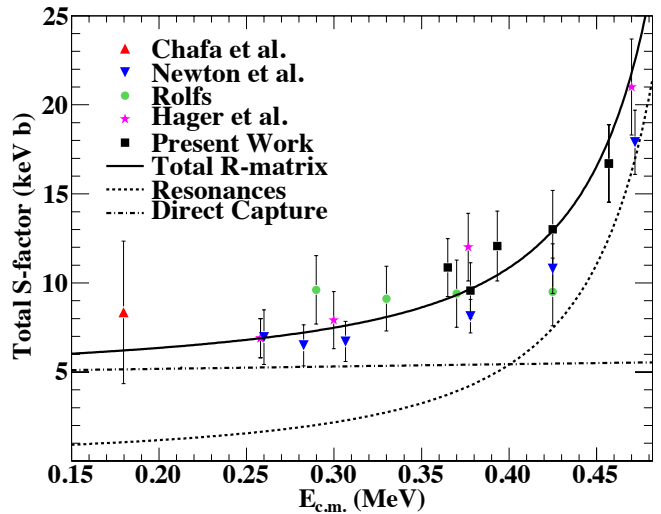}
\caption{(Color online) Comparison of the $S$-factor obtained from the $R$-matrix fit of the present data set with literature experimental data. The thick continuous line represents the extrapolation of the total $S$-factor from the best $R$-matrix fit, with background poles included and some ANC values fixed (``Total $R$-matrix''). The dashed-dotted line and the thick dashed line are the contributions of the direct capture (background poles included) and the high energy resonances respectively. All data shown include total uncertainties.}
\label{FIG:comparison}
\end{figure}

In addition, to test the dependence of the extrapolation on the adopted fixed ANC values, the latter were varied upwards and downwards and the results of the fits were compared with each other. For the R/DC$\rightarrow$937 transition, a $23\%$ variation of the ANC resulted in a change in the zero energy extrapolation by $15\%$. Note that a $23\%$ variation of the ANC corresponds to a $50\%$ variation in the respective spectroscopic factor. The reported uncertainties of the spectroscopic factors in~\cite{polsky1969} are $25\%$, but given the uncertainties involved in converting spectroscopic factors from transfer reactions to ANCs, a conservative $50\%$ uncertainty was adopted. All the tests mentioned allow for a reasonable upper limit estimation of the uncertainty attributed solely to the $R$-matrix extrapolation, by adding the variations in quadrature. A $10\%$ uncertainty was also added to the extrapolated zero energy total S-factor, related to how well the data constrain the free parameters. The result for the uncertainty of the extrapolation of the total $S$-factor to $E=0$ is $\sim19\%$, in addition to the experimental uncertainty which amounts to $\sim11\%$.

In Fig.~\ref{FIG:comparison}, the results from this study are compared with recent experimental data. The energy dependence of the $R$-matrix extrapolation agrees very well with the low-energy data by Newton \textit{et al.}~\cite{newton2010} (blue inverse triangles) and Hager \textit{et al.}~\cite{hager2012} (purple stars). In the case of the Hager data the agreement with the present extrapolation is remarkable. The deviation from the Newton data is approximately $15\%$ and can be attributed to the uncertainties of the two experiments ($\sim15\%$ for the Newton data and $11\%$ for the present data). On the other hand the energy dependence of the data from Rolfs\footnote{The direct capture data by Rolfs are not subject to the normalization factor that the resonance strengths were corrected by, since they were measured relative to $^{16}$O$(p,\gamma)^{17}$F, as discussed in section~\ref{SEC:results}.} (green circles) disagrees with the present extrapolation, given that the error bars show the total uncertainty, and an overall normalization factor would not change the shape of the data. No conclusion can be drawn by the comparison of the present extrapolation and the Chafa \textit{et al.} point (red square) given its large uncertainty. It should be emphasized that the literature low-energy data were not included in the fit, and thus the extrapolation depends solely on the present work's data set and on the experimental ANCs from the transfer reactions. 

 \begin{table}[t]
 \caption{S(0) values of $^{17}$O$(p,\gamma)^{18}$F for each measured $\gamma$-ray transition \label{T:sfactorcontributions}}
 \begin{ruledtabular}
 \begin{tabular}{lc}
      Transition (keV) & $S(0)$ (keV b)~ \\
\hline
       R/DC$\rightarrow$937	& $1.7\pm 0.3$~ \\
       R/DC$\rightarrow$1121	& $0.66\pm 0.13$~ \\
       R/DC$\rightarrow$1700	& $0.013\pm 0.002$~ \\
       R/DC$\rightarrow$2523	& $0.17\pm 0.03$~ \\
       R/DC$\rightarrow$3062	& $0.66\pm 0.10$~ \\
       R/DC$\rightarrow$3791	& $0.032\pm 0.005$~ \\
       R/DC$\rightarrow$3839	& $0.93\pm 0.14$~ \\
       R/DC$\rightarrow$4115	& $0.55\pm 0.08$~ \\
       R/DC$\rightarrow$4652	& $0.21\pm 0.03$~ \\
       R/DC$\rightarrow$4964	& $0.49\pm 0.07$~ \\
\end{tabular}
 \end{ruledtabular}
 \end{table}
Table~\ref{T:sfactorcontributions} lists the calculated contributions of all the measured transitions to the total $S$-factor at zero energy. The uncertainties come from the estimated $15\%$ uncertainty that arises from the choice of the ANC values. The experimental uncertainty is not included.

 \begin{table}[b]
 \caption{$S(0)$ values of $^{17}$O$(p,\gamma)^{18}$F from literature \label{T:sfactors}}
 \begin{ruledtabular}
 \begin{tabular}{lc}
      Source & $S(0)$ (keV b) \\
\hline
      Rolfs \cite{rolfs1973} & $9.4\pm 1.9$~ \\
      Fox \textit{et al.} \cite{fox2005} &$4.2\pm 1.9$\footnote[1]{The listed $S(0)$ value of Fox \emph{et al.}~\cite{fox2005} was calculated by adding a resonance contribution of $S_{res}=0.52$ keV b (obtained from the present study) to the reported DC value $3.7$ keV b.}  \\
      Chafa \textit{et al.} \cite{chafa2007} &$6.2\pm 3.1$~  \\
      Newton \textit{et al.} \cite{newton2010} & $5.1\pm 0.9$\footnote[2]{The Newton and Hager values were estimated by normalizing the present resonant and direct capture contributions to the respective experimental data.} \\
      Hager \textit{et al.} \cite{hager2012} & $5.8\pm 0.9$\footnotemark[2] \\
      Present work & $5.4\pm \text{(th.)} 1.0 \pm \text{(exp.)} 0.6$~ \\
\end{tabular}
 \end{ruledtabular}
 \end{table}
In Table~\ref{T:sfactors}, the present $S(0)$ calculation is compared with literature values. The uncertainty of the present $S(0)$ was calculated from the uncertainties of the experimental procedure and the uncertainty related to the theoretical fit. The $S(0)$ values for Newton et al.~\cite{newton2010} and Hager et al.~\cite{hager2012} were calculated by fitting the shape of the resonant and direct capture contributions as calculated from this work to their experiment data, with the two parameters being the normalization of each contribution ($S_{tot}=a_1 S_{res}+a_2 S_{DC}$). The value obtained here for the data of Newton et al. ($5.1\pm0.9$ keV b) is identical to the one reported in~\cite{newton2010} ($5.1\pm1.1$ keV b), but with a smaller uncertainty given that part of the previous uncertainty came from the resonant contribution. No attempt was made by Hager \textit{et al.}~\cite{hager2012} to extrapolate the $S$-factor to $E=0$ from their data. 

Good agreement is observed with all previous results except for the value proposed by Rolfs, which disagrees with the present result by a factor of $1.6$. The reason for this discrepancy is not clear. The evidence given in Sec.~\ref{SEC:results} of this paper for a $40\%$ overall normalization error in the Rolfs data (not to be confused with the $0.62$ normalization factor applied to the resonance strengths reported by Rolfs~\cite{rolfs1973}) could not solve the issue, as the energy dependence of the data is different. For this reason a definite conclusion cannot be reached. 

The direct capture calculations by Fox \textit{et al.}~\cite{fox2005} were performed with a Wood-Saxon bound state potential and hard sphere phase shifts for the scattering wave functions, and relied on measured spectroscopic factors. The result of that calculation was an $S$-factor with a slope very close to zero. The present extrapolation agrees with these results as well, at least for the low-energy region of astrophysical interest. The differences can be traced to how the calculations were normalized. 

Given the consistency between the experimental data by Newton \textit{et al.}, Hager \textit{et al.}~\cite{newton2010,hager2012}, and this work, and the relative agreement in the shape of the direct capture $S$-factor between Fox \textit{et al.} and this work, it is recommended that the direct capture contribution be taken as the weighted average of the results from the calculated $S_{DC}$'s from the Newton ($4.7\pm0.7$ keV b) and Hager ($5.3\pm0.8$ keV b) data and this work ($4.9\pm1.1$ keV b)\footnote{Here the theoretical and experimental uncertainties of the direct capture part were assumed to add in quadrature.}. The result of the weighted average is $S_{DC}=4.8\pm0.5$ keV b.

\section{REACTION RATES \label{SEC:rates}}

The total thermonuclear rate for the $^{17}$O$(p,\gamma)^{18}$F reaction was calculated by direct numerical integration of the formula
\begin{eqnarray}
	 N_A \left<\sigma v \right> = 3.7318\times10^{10} \mu^{-1/2} T_9^{-3/2} \nonumber \\
	 						\times \int_{0}^{\infty} \sigma(E) E e^{-11.605 E/T_9} dE,  \label{reactionrate}
\end{eqnarray}
where the rate is in units of cm$^3$ s$^{-1}$ mole$^{-1}$, $T_9$ is the stellar temperature in GK, $\mu$ is the reduced mass, $E$ is the center-of-mass energy in MeV and $\sigma(E)$ is the reaction cross section in barns. However, numerical integration is less reliable for narrow resonances where sharp changes of the cross section occur. The reaction rate in these cases was calculated using the formula for narrow and isolated resonances
\begin{eqnarray}
	 N_A \left<\sigma v \right>_r = 1.540\times10^{11} (\mu T_9)^{-3/2} (\omega \gamma) \nonumber \\
	 						\times e^{-11.605 E_{r}/T_9},  \label{narrowrate}
\end{eqnarray}
where $\omega\gamma$ and $E_{r}$ are the resonance strength and resonance energy in the center-of-mass in MeV. The strengths for the two low-energy resonances at $E_{r}=65.1$ and $183$ keV were taken from Fox \textit{et al.}~\cite{fox2005}, whereas their energies were taken from Chafa \textit{et al.}~\cite{chafa2007}, following the suggestion of the reaction rate evaluation by Iliadis \textit{et al.}~\cite{iliadis2010}, which is the recommended evaluation by {\sc REACLIB} \cite{REACLIB} for this reaction. The direct capture reaction contribution  was taken to be the weighted average of the results from the analysis of the present experimental data, Hager \textit{et al.}, and Newton \textit{et al.}. The direct capture contribution in the {\sc REACLIB} evaluation was adopted from the calculation by Newton \textit{et al.}. Interference effects between the narrow resonances and the direct capture were neglected, as the mechanisms mainly proceed through different orbital angular momenta or different $\gamma$-ray multipolarities than these resonances.
\begin{figure}[t]
\centering
\includegraphics[width=0.45 \textwidth]{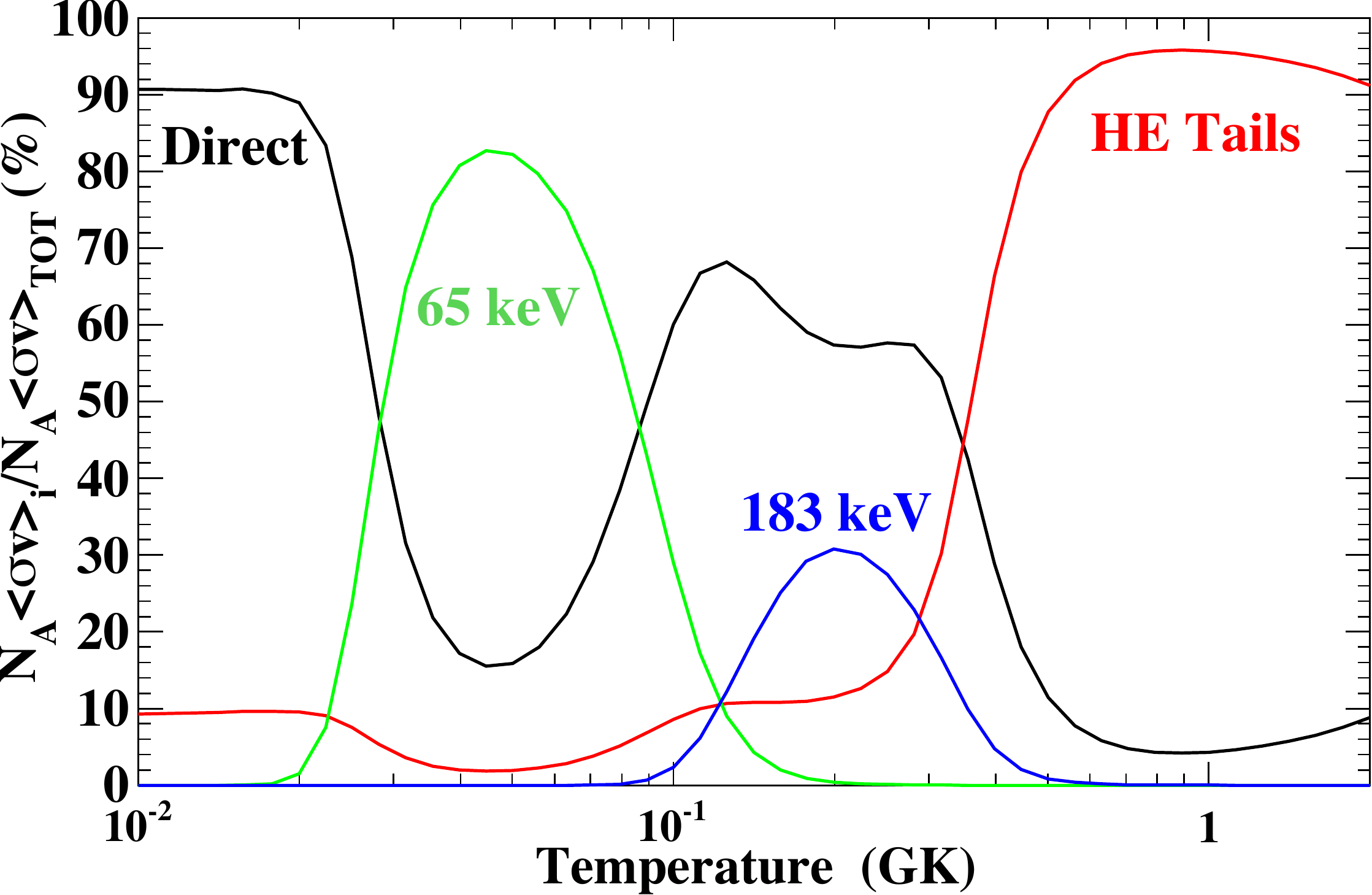}
\caption[Contributions to the total reaction rate.]{(Color online) Contributions of different components to the total reaction rate as a function of temperature. HE Tails stands for ``High Energy Tails'' contributions. The importance of the direct capture component is evident.}
\label{FIG:reactionrate_perc}
\end{figure}

Figure~\ref{FIG:reactionrate_perc} shows the percentage contributions of different reaction components to the total reaction rate as a function of temperature. ``HE Tails'' includes the contributions of the two narrow resonances at $E_{r}=490$ and $531$ keV and more importantly the two broad resonances at $E_{r}=557$ and $677$ keV. The importance of the direct capture contribution to the total reaction rate is evident.

\begin{figure}[t]
\centering
\includegraphics[width=0.45 \textwidth]{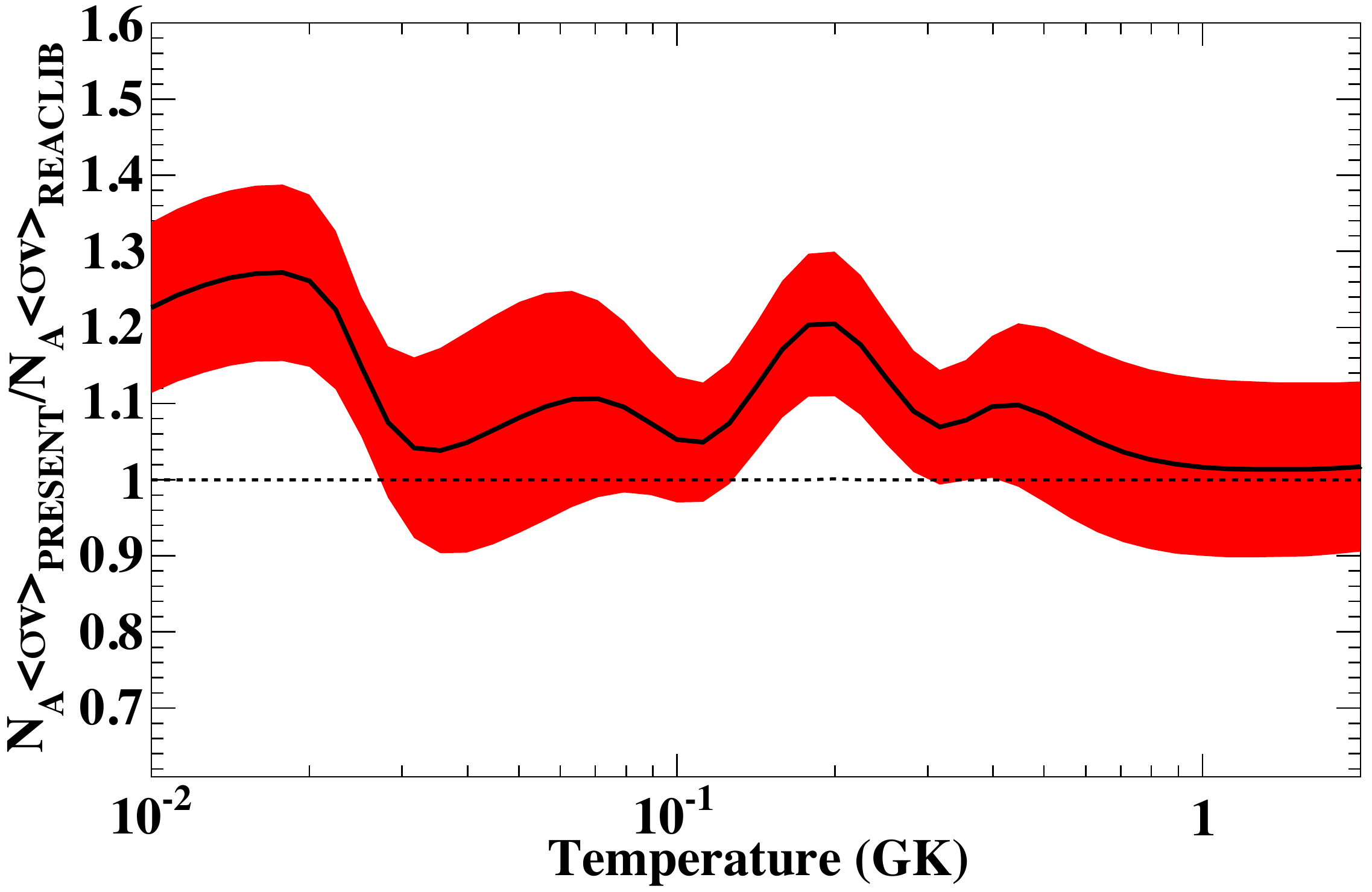}
\caption{Reaction rate ratio of the present reaction rate to the {\sc REACLIB} rate as a function of temperature. The uncertainty band includes the estimated uncertainties of the direct capture and resonant components.}
\label{FIG:reactionrate_ratio}
\end{figure}

The total reaction rates obtained from this work are tabulated in Table~\ref{T:rates}, in the appendix. The ratio of the present reaction rate to the {\sc REACLIB} rate as a function of temperature is shown in Fig.~\ref{FIG:reactionrate_ratio}. The uncertainty band includes the uncertainties for the direct capture and the high energy resonance tails, as were estimated from this work, and the uncertainties of the two narrow resonances at $E_{r}=65.1$ and $183$ keV, as reported in~\cite{fox2005} and ~\cite{chafa2007} respectively. At the temperature range $T=0.03-0.1$ GK relevant for equilibrium hydrogen burning (red giants, AGB stars), the reaction rate calculated from this work is higher by up to $20\%$. The reason for this is not clear given the fact that the direct capture component used by~\cite{iliadis2010} is only by $4\%$ lower than that used for this calculation. Nevertheless, there is still reasonable agreement considering the uncertainties. At the temperatures relevant for classical novae ($T=0.1-0.4$ GK) the present reaction rate is also higher than the recommended rate from {\sc REACLIB} by up to $20\%$. 

\section{SUMMARY AND CONCLUSION \label{SEC:conclusion}}

The excitation functions of the $^{17}$O$(p,\gamma)^{18}$F has been measured in the energy range of $E_{p}= 365$ keV to $1800$ keV at angles of $\theta_{lab}=0^\circ,45^\circ,90^\circ$, and $135^\circ$ for 10 different primary transitions to states in $^{18}$F. New resonance strengths and branching ratios were obtained for the observed resonances in this energy range. An $R$-matrix analysis was performed to extrapolate the astrophysical $S$-factor to lower energies relevant for several astrophysical scenarios. In this analysis all primary transitions observed in this experiment were simultaneously fitted together with the $^{17}$O$(p,\alpha)^{14}$N data of reference~\cite{kieser1979}. The resulting total S-factor is dominated by the DC contributions and agrees with recent low-energy measurements~\cite{newton2010,hager2012} within the experimental uncertainties.

The reaction rate deduced from the $R$-matrix analysis is slightly higher (up to $20\%$) than the rates in the compilations of reference~\cite{iliadis2010} but in agreement within the respective uncertainties. The main uncertainty of the reaction rate is the direct result of the uncertainty of the DC contribution. In the present experimental energy range the cross section is dominated by broad and strong resonances masking the DC cross sections except for the highest energies. The uncertainty could be further reduced by extending the literature data to lower energies ($<260$ keV). Such measurements are challenging and require a low background environment such as an underground laboratory.

\begin{acknowledgments}
The authors would like to thank the technical stuff of the Nuclear Science Laboratory, at the University of Notre Dame. This work was funded in part by the National Science Foundation through grant number Phys-0758100 and the Joint Institute for Nuclear Astrophysics grant number Phys-0822648.
\end{acknowledgments}

\appendix

\section{$R$-matrix fits}
\begin{figure*}[h]
\centering
\includegraphics[width=1.0 \textwidth]{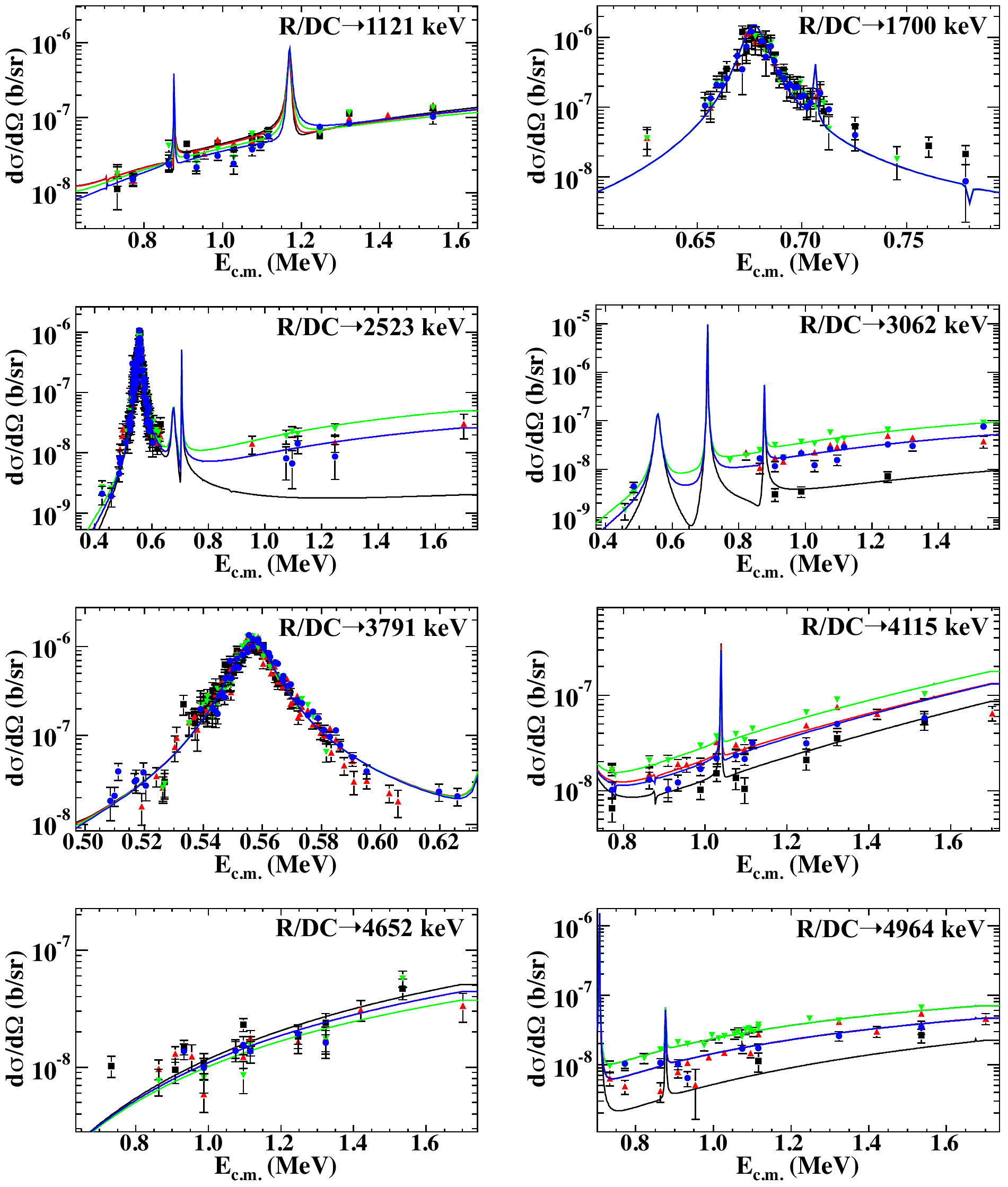}
\caption{(Color online) $R$-matrix fits of the differential cross section of $^{17}$O$(p,\gamma)^{18}$F. Each plot corresponds to a primary $\gamma$-ray transition to a different bound state of $^{18}$F. The black squares correspond to $0^\circ$, red triangles to $45^\circ$, green inverse triangles to $90^\circ$, and blue circles to $135^\circ$. The lines are color coded the same as the data points. Only statistical errors are displayed.}
\label{FIG:plots8}
\end{figure*}

\begin{figure*}[h]
\centering
\includegraphics[width=0.58 \textwidth]{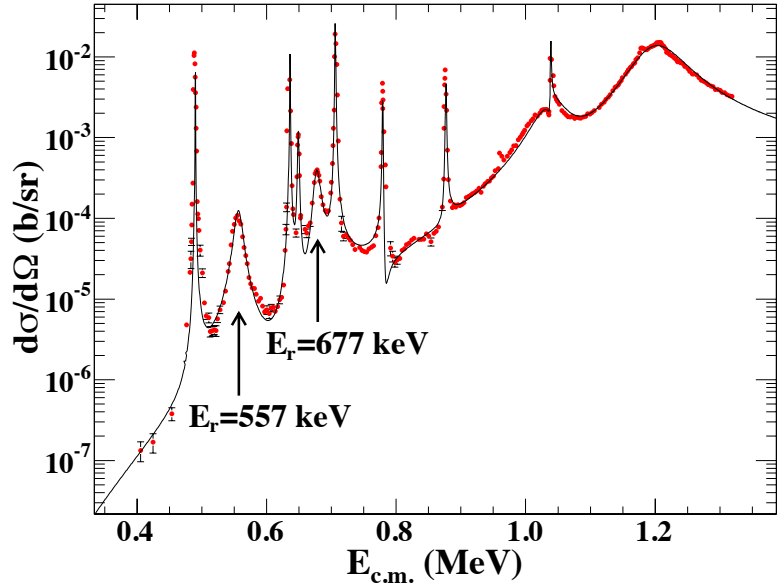}
\caption{(Color online) $R$-matrix fit of the $^{17}$O$(p,\alpha)^{14}$N Kieser data~\cite{kieser1979} obtained at $135^\circ$. The arrows point to the two resonances of interest for the $^{17}$O$(p,\gamma)^{18}$F case. Target effects for this channel were not taken into account. For this reason the cross section scale does not apply to the data points near the maxima of the narrow resonances. This simplification has no effect on the results for the broad resonances (see text for details). Only statistical errors are displayed.}
\label{FIG:kieser}
\end{figure*}

All the remaining $R$-matrix fits are shown in Fig.~\ref{FIG:plots8} and ~\ref{FIG:kieser}. Some resonances in energy regions that were not covered by the experiment, were also included, when they could influence regions where data were taken. All fits shown in this appendix and in the main text were performed simultaneously. 

\section{Recommended reaction rates}
The reaction rates as a function of temperature calculated in this work are tabulated in table~\ref{T:rates}. The upper and lower rates correspond to the $1\sigma$ uncertainty band shown in Fig.~\ref{FIG:reactionrate_ratio} in the main text.

\begin{table*}[b]
\caption{$^{17}$O$(p,\gamma)^{18}$F reaction rates \label{T:rates}}
\begin{ruledtabular}
\begin{tabular}{cccccccc}
      \multicolumn{1}{c}{$T_9$} 	&  \multicolumn{1}{c}{Lower Rate}  &  \multicolumn{1}{c}{Reaction Rate} & \multicolumn{1}{c}{Upper Rate} & \multicolumn{1}{|c}{$T_9$} 	&  \multicolumn{1}{c}{Lower Rate}  &  \multicolumn{1}{c}{Reaction Rate} & \multicolumn{1}{c}{Upper Rate} \\ 
      \hline
$	0.010	$ & $	4.14\times10^{-25}	$ & $	4.55\times10^{-25}	$ & $	4.97\times10^{-25}	$ & \multicolumn{1}{|c}{	$	0.224	$} & $	4.35\times10^{-04}	$ & $	4.72\times10^{-04}	$ & $	5.09\times10^{-04}	$ \\
$	0.011	$ & $	6.77\times10^{-24}	$ & $	7.45\times10^{-24}	$ & $	8.13\times10^{-24}	$ & \multicolumn{1}{|c}{	$	0.251	$} & $	1.13\times10^{-03}	$ & $	1.22\times10^{-03}	$ & $	1.32\times10^{-03}	$ \\
$	0.013	$ & $	1.10\times10^{-22}	$ & $	1.22\times10^{-22}	$ & $	1.33\times10^{-22}	$ & \multicolumn{1}{|c}{	$	0.282	$} & $	2.87\times10^{-03}	$ & $	3.10\times10^{-03}	$ & $	3.33\times10^{-03}	$ \\
$	0.014	$ & $	1.43\times10^{-21}	$ & $	1.58\times10^{-21}	$ & $	1.72\times10^{-21}	$ & \multicolumn{1}{|c}{	$	0.316	$} & $	7.57\times10^{-03}	$ & $	8.14\times10^{-03}	$ & $	8.72\times10^{-03}	$ \\
$	0.016	$ & $	1.74\times10^{-20}	$ & $	1.91\times10^{-20}	$ & $	2.09\times10^{-20}	$ & \multicolumn{1}{|c}{	$	0.355	$} & $	2.22\times10^{-02}	$ & $	2.40\times10^{-02}	$ & $	2.57\times10^{-02}	$ \\
$	0.018	$ & $	2.15\times10^{-19}	$ & $	2.36\times10^{-19}	$ & $	2.58\times10^{-19}	$ & \multicolumn{1}{|c}{	$	0.398	$} & $	7.34\times10^{-02}	$ & $	8.03\times10^{-02}	$ & $	8.71\times10^{-02}	$ \\
$	0.020	$ & $	2.30\times10^{-18}	$ & $	2.53\times10^{-18}	$ & $	2.75\times10^{-18}	$ & \multicolumn{1}{|c}{	$	0.447	$} & $	2.56\times10^{-01}	$ & $	2.84\times10^{-01}	$ & $	3.12\times10^{-01}	$ \\
$	0.022	$ & $	2.23\times10^{-17}	$ & $	2.44\times10^{-17}	$ & $	2.65\times10^{-17}	$ & \multicolumn{1}{|c}{	$	0.501	$} & $	8.65\times10^{-01}	$ & $	9.68\times10^{-01}	$ & $	1.07\times10^{+00}	$ \\
$	0.025	$ & $	2.28\times10^{-16}	$ & $	2.48\times10^{-16}	$ & $	2.68\times10^{-16}	$ & \multicolumn{1}{|c}{	$	0.562	$} & $	2.68\times10^{+00}	$ & $	3.01\times10^{+00}	$ & $	3.35\times10^{+00}	$ \\
$	0.028	$ & $	2.62\times10^{-15}	$ & $	2.88\times10^{-15}	$ & $	3.15\times10^{-15}	$ & \multicolumn{1}{|c}{	$	0.631	$} & $	7.49\times10^{+00}	$ & $	8.44\times10^{+00}	$ & $	9.40\times10^{+00}	$ \\
$	0.032	$ & $	2.77\times10^{-14}	$ & $	3.13\times10^{-14}	$ & $	3.49\times10^{-14}	$ & \multicolumn{1}{|c}{	$	0.708	$} & $	1.88\times10^{+01}	$ & $	2.12\times10^{+01}	$ & $	2.36\times10^{+01}	$ \\
$	0.036	$ & $	2.71\times10^{-13}	$ & $	3.11\times10^{-13}	$ & $	3.52\times10^{-13}	$ & \multicolumn{1}{|c}{	$	0.794	$} & $	4.25\times10^{+01}	$ & $	4.80\times10^{+01}	$ & $	5.35\times10^{+01}	$ \\
$	0.040	$ & $	2.11\times10^{-12}	$ & $	2.45\times10^{-12}	$ & $	2.79\times10^{-12}	$ & \multicolumn{1}{|c}{	$	0.891	$} & $	8.74\times10^{+01}	$ & $	9.88\times10^{+01}	$ & $	1.10\times10^{+02}	$ \\
$	0.045	$ & $	1.38\times10^{-11}	$ & $	1.61\times10^{-11}	$ & $	1.84\times10^{-11}	$ & \multicolumn{1}{|c}{	$	1.000	$} & $	1.65\times10^{+02}	$ & $	1.86\times10^{+02}	$ & $	2.08\times10^{+02}	$ \\
$	0.050	$ & $	7.24\times10^{-11}	$ & $	8.43\times10^{-11}	$ & $	9.62\times10^{-11}	$ & \multicolumn{1}{|c}{	$	1.122	$} & $	2.87\times10^{+02}	$ & $	3.24\times10^{+02}	$ & $	3.62\times10^{+02}	$ \\
$	0.056	$ & $	3.25\times10^{-10}	$ & $	3.76\times10^{-10}	$ & $	4.27\times10^{-10}	$ & \multicolumn{1}{|c}{	$	1.259	$} & $	4.66\times10^{+02}	$ & $	5.26\times10^{+02}	$ & $	5.86\times10^{+02}	$ \\
$	0.063	$ & $	1.28\times10^{-09}	$ & $	1.46\times10^{-09}	$ & $	1.65\times10^{-09}	$ & \multicolumn{1}{|c}{	$	1.413	$} & $	7.10\times10^{+02}	$ & $	8.01\times10^{+02}	$ & $	8.92\times10^{+02}	$ \\
$	0.071	$ & $	4.46\times10^{-09}	$ & $	5.05\times10^{-09}	$ & $	5.65\times10^{-09}	$ & \multicolumn{1}{|c}{	$	1.585	$} & $	1.02\times10^{+03}	$ & $	1.15\times10^{+03}	$ & $	1.28\times10^{+03}	$ \\
$	0.079	$ & $	1.44\times10^{-08}	$ & $	1.61\times10^{-08}	$ & $	1.78\times10^{-08}	$ & \multicolumn{1}{|c}{	$	1.778	$} & $	1.39\times10^{+03}	$ & $	1.57\times10^{+03}	$ & $	1.74\times10^{+03}	$ \\
$	0.100	$ & $	1.45\times10^{-07}	$ & $	1.57\times10^{-07}	$ & $	1.69\times10^{-07}	$ & \multicolumn{1}{|c}{	$	1.995	$} & $	1.82\times10^{+03}	$ & $	2.05\times10^{+03}	$ & $	2.27\times10^{+03}	$ \\
$	0.112	$ & $	4.70\times10^{-07}	$ & $	5.08\times10^{-07}	$ & $	5.46\times10^{-07}	$ & \multicolumn{1}{|c}{	$	2.239	$} & $	2.29\times10^{+03}	$ & $	2.56\times10^{+03}	$ & $	2.84\times10^{+03}	$ \\
$	0.126	$ & $	1.57\times10^{-06}	$ & $	1.70\times10^{-06}	$ & $	1.82\times10^{-06}	$ & \multicolumn{1}{|c}{	$	2.512	$} & $	2.77\times10^{+03}	$ & $	3.10\times10^{+03}	$ & $	3.43\times10^{+03}	$ \\
$	0.141	$ & $	5.31\times10^{-06}	$ & $	5.74\times10^{-06}	$ & $	6.17\times10^{-06}	$ & \multicolumn{1}{|c}{	$	2.818	$} & $	3.27\times10^{+03}	$ & $	3.65\times10^{+03}	$ & $	4.02\times10^{+03}	$ \\
$	0.159	$ & $	1.74\times10^{-05}	$ & $	1.88\times10^{-05}	$ & $	2.03\times10^{-05}	$ & \multicolumn{1}{|c}{	$	3.162	$} & $	3.77\times10^{+03}	$ & $	4.19\times10^{+03}	$ & $	4.61\times10^{+03}	$ \\
$	0.178	$ & $	5.41\times10^{-05}	$ & $	5.87\times10^{-05}	$ & $	6.33\times10^{-05}	$ & \multicolumn{1}{|c}{	$	3.548	$} & $	4.28\times10^{+03}	$ & $	4.74\times10^{+03}	$ & $	5.19\times10^{+03}	$ \\
$	0.200	$ & $	1.58\times10^{-04}	$ & $	1.72\times10^{-04}	$ & $	1.85\times10^{-04}	$ & \multicolumn{1}{|c}{	$	3.981	$} & $	4.81\times10^{+03}	$ & $	5.30\times10^{+03}	$ & $	5.78\times10^{+03}	$ \\
\end{tabular}
\end{ruledtabular}
\end{table*}


\begin{thebibliography}{20}
\bibitem{iliadisstars}
	C. Iliadis, \emph{Nuclear Physics of Stars} (Wiley-VCH, Weinheim, 2007).
\bibitem{harris1988}
	M. J. Harris, D. L. Lambert, and V. V. Smith, \emph{Astrophys. J.} {\bf325}, 768 (1988).
\bibitem{dearborn1992}
	D. S. P. Dearborn, \emph{Phys. Rep.} {\bf210}, 367 (1992).
\bibitem{nollett2003}
	K. M. Nollett, M. Busso, and G. J. Wasserburg, \emph{Astrophys. J.} {\bf582}, 1036 (2003).
\bibitem{abia2011}
	C. Abia, K. Cunha, S. Cristallo, P. de Laverny, I. Dom'nguez, A. Recio-Blanco, V. V. Smith, and O. Straniero, \emph{Astrophys. J.} {\bf 737}, L8 (2011).
\bibitem{palmerini2011}
	S. Palmerini, M. La Cognata, S. Cristallo, and M. Busso, \emph{Astrophys. J.} {\bf 729}, 3 (2011).
\bibitem{novae}
	S. Starrfield, C. Iliadis, and W. R. Hix, in \emph{Thermonuclear Processes, Classical Novae} (2nd ed.) edited by M. F. Bode and A. Evans (Cambridge University Press, Cambridge, England, 2008).
\bibitem{iliadis2002}
	C. Iliadis, A. Champagne, J. Jose, S. Starrfield and P. Tupper, \emph{Astrophys. J. Suppl.} {\bf142}, 105 (2002).
\bibitem{fox2004}
	C.~Fox, C.~Iliadis, A.~E.~Champagne, A.~Coc, J.~Jos\'e, R.~Longland, J.~Newton, J.~Pollanen and R.~Runkle, \emph{Phys. Rev. Lett.} {\bf93}, 081102 (2004).	
\bibitem{integral}
	J. G\'omez-Gomar, M. Hernanz, J. Jos\'e, and J. Isern, \emph{Mon. Not. R. Astron. Soc.} {\bf296}, 913 (1998).
\bibitem{fox2005}
	C.~Fox, C.~Iliadis, A.~E.~Champagne, R.~P.~Fitzgerald, R.~Longland, J.~Newton, J.~Pollanen, R.~Runkle, \emph{Phys. Rev. C} {\bf71}, 055801 (2005).
\bibitem{chafa2007}
	A. Chafa \textit{et al.}, \emph{Phys. Rev. C} {\bf75}, 035810 (2007).
\bibitem{rolfs1973}
	C. Rolfs, \emph{Nucl. Phys. A} {\bf217}, 29 (1973).
\bibitem{newton2010}
	J. R. Newton \textit{et al.}, \emph{Phys. Rev. C} {\bf81}, 045801 (2010).
\bibitem{hager2012}
	U. Hager \textit{et al.}, \emph{Phys. Rev. C} {\bf85}, 035803 (2012).
\bibitem{azure}
	R.~E.~Azuma and E.~Uberseder \textit{et al.}, \emph{Phys. Rev. C} {\bf81}, 045805 (2010).
\bibitem{aluminum}
	A. Anttila, J. Keinonen, M. Hautala, I. Forsblom, \emph{Nucl. Instrum. Methods} {\bf147}, 501 (1977).
\bibitem{imbriani2005}
	G. Imbriani et al., \emph{Eur. Phys. J. A} {\bf25}, 455 (2005).
\bibitem{efficiency1998}
	Z. Kis, B. Fazekas, J. \"Ost\"or, Zs. R\'evay, T. Belgya, G.L. Moln\'ar, L. Koltay, \emph{Nucl. Instrum. Methods A} {\bf418}, 374 (1998).
\bibitem{geant4}
	Geant4 Collaboration. \emph{Nucl. Instrum. Methods A} {\bf506}, 250 (2003).
\bibitem{rose1953}
	M. E. Rose, \emph{Phys. Rev.} {\bf91}, 610 (1953).
\bibitem{anodization}
	D. Phillips and J. P. S. Pringle, \emph{Nucl. Instrum. Methods} {\bf135}, 389 (1976).
\bibitem{srim}
	J. F. Ziegler, J. P. Biersack, and M. D. Ziegler, SRIM: The Stopping and Range of Ions in Matter (2008) [www.srim.org].
\bibitem{tilley1995}
	D.R. Tilley, H.R. Weller, C.M. Cheves, R.M. Chasteler, \emph{Nucl. Phys. A} {\bf595}, 1 (1995).
\bibitem{iliadis2001}
	C. Iliadis, J. DÕAuria, S. Starrfield, W. J. Thompson, and M. Wiescher, \emph{Astrophys. J. Suppl.} {\bf134}, 151 (2001).
\bibitem{rolfs1973b}
	C. Rolfs, \emph{Nucl. Phys. A} {\bf199}, 257 (1973).
\bibitem{sens1973}
	J. C. Sens, A. Pape, R. Armbruster, \emph{Nucl. Phys. A} {\bf199}, 241 (1973).
\bibitem{kieser1979}
	W. E. Kieser, R. E. Azuma and K. P. Jackson, \emph{Nucl. Phys.} {\bf A331}, 155 (1979).
\bibitem{polsky1969}
	L. M. Polsky \textit{et al.}, \emph{Phys. Rev.} {\bf186}, 966 (1969).
\bibitem{landre1989}
	V. Landre \textit{et al.}, \emph{Phys. Rev. C} {\bf40}, 1972 (1989).

\bibitem{iliadis2010}
	C. Iliadis, R. Longland, A.E. Champagne, A. Coc, R. Fitzgerald, \emph{Nucl. Phys. A}, {\bf841}, 31 (2010).
\bibitem{REACLIB}
 	R. H. Cyburt \textit{et al.}, \emph{Astrophys. J. Suppl.}, 189:240 (2010) [http://groups.nscl.msu.edu/jina/reaclib/db/index.php].


\end{thebibliography}

\end{document}